\documentclass[a4paper,12pt]{article}
\pdfoutput=1

\usepackage{jheppub}
\usepackage{bm,latexsym,amsmath,amssymb,amsfonts,mathtools,mathrsfs}
\usepackage{orcidlink}

\usepackage{graphicx}
\usepackage{wrapfig}

\usepackage{color}
\usepackage{subcaption}

\definecolor{nicered}{rgb}{0.7,0.1,0.1}
\definecolor{nicegreen}{rgb}{0.1,0.5,0.1}

\usepackage{todonotes}
\definecolor{palatinate}{RGB}{128, 49, 123}

\def\cee{{\relax\hbox{$\inbar\kern-.3em{\rm C}$}}}

\newcommand{\be}{\begin{equation}}
\newcommand{\ee}{\end{equation}}
\newcommand{\bea}{\begin{eqnarray}}
\newcommand{\eea}{\end{eqnarray}}
\newcommand{\beal}{\begin{aligned}}
\newcommand{\eeal}{\end{aligned}}
\newcommand{\bec}{\begin{cases}}
\newcommand{\eec}{\end{cases}}

\newcommand{\mrm}[1]{\mathrm{#1}}
\newcommand{\mc}[1]{\mathcal{#1}}

\newcommand{\lr}[1]{\!\left(#1\right)}

\newcommand{\lrb}[1]{\!\left[#1\right]}

\newcommand{\vb}[1]{{\bf #1}}

\newcommand{\RR}{\mathbb{{R}}}

\title{Acceleration and Rotation in Three Dimensions}

\author[a,b]{Cameron R D Bunney\orcidlink{0000-0003-3998-4716}}
\author[c,d]{Ruth Gregory\orcidlink{0000-0003-0424-3440}}
\author[d,e]{Robert B Mann\orcidlink{0000-0002-5859-2227}}

\affiliation[a]{School of Physics and Astronomy, University of Nottingham, University Park, Nottingham, NG7 2RD, UK}
\affiliation[b]{Nottingham Centre of Gravity, University of Nottingham, University Park, Nottingham, NG7 2RD, UK}
\affiliation[c]{Department of Physics, King’s College London, University of London, Strand, London, WC2R 2LS, UK}
\affiliation[d]{Perimeter Institute for Theoretical Physics, $31$ Caroline St., Waterloo, Ontario, N2L 2Y5, Canada}
\affiliation[e]{Department of Physics and Astronomy, University of Waterloo, Waterloo, Ontario, N2L 3G1, Canada}

\emailAdd{cameron.r.d.bunney@gmail.com}
\emailAdd{ruth.gregory@kcl.ac.uk}
\emailAdd{rbmann@uwaterloo.ca}

\abstract{
We present a new family of exact solutions
representing accelerating, {\it rotating}, point particle and black hole 
solutions in three-dimensional Einstein gravity. 
We investigate their geometric properties and the global
embedding in three dimensional anti-de Sitter space.
Our results extend the catalogue of exact solutions in three-dimensional Einstein 
gravity and provide a useful setting for the study of black hole thermodynamics, 
asymptotic symmetries, and holography.
}

\begin{document}

\maketitle

\section{Introduction}

Three dimensional (3D) anti-de Sitter (AdS) spacetime provides a useful theoretical laboratory for investigating fundamental aspects of classical and quantum gravity. Despite the absence of local gravitational degrees of freedom, the ability to perform non-trivial identifications in the geometry gives rise to interesting physically relevant global structures, exemplified by the black hole solutions discovered by Bañados, Teitelboim, and Zanelli (BTZ)  \cite{BTZoriginal,BTZgeometry}. These geometries have played a central role in the study of holography, thermodynamics, and quantum gravity, indeed, many features of higher-dimensional black holes are mirrored in this considerably simpler setting.

In 4D, an exact solution representing an accelerating black hole, the C-metric, has been known for more than a century \cite{LeviCivita:1917,Weyl:1919}, although its physical interpretation and relevance were not fully explored until much later 
\cite{KinnersleyWalker:1970,Plebanski:1976gy,Podolsky:2002nk}. The C-metric is an axially symmetric static (or stationary) metric that is characterised by an asymmetry in conical singularities emanating from the poles of a compact horizon. The conical singularities are conventionally interpreted as cosmic strings or struts, and the asymmetry of the deficit/excess angle provides a force imbalance due to the action of the differing tensions on the black hole horizon. The conical singularity can be smoothed out locally by introducing a source for the cosmic string \cite{Achucarro:1995nu,Gregory:1995hd,Gregory:2013xca}, thus the C-metric can be understood as an effective description of a non-singular
geometry. The C-metrics (or closely related geometries) have a range of physically interesting applications, from black hole pair creation 
\cite{Dowker:1993bt,Emparan:1995je,Gregory:1995hd,Hawking:1995zn,MannRoss:1995} to thermodynamics and holography 
\cite{Appels:2016uha,Appels:2017xoe,Anabalon:2018ydc,Anabalon:2018qfv,Gregory:2019dtq,CisternaMannHairyBTZ,Xue:2025ajm}. When rotation is incorporated \cite{Bicak:1999sa,Hong:2004dm}, the resulting solutions exhibit an even richer interplay between acceleration horizons, ergoregions, and global causal structure, however, while rotating accelerating solutions in 4D have been extensively investigated, to our knowledge there is no known analog in 3D.

Various special cases of accelerating solutions in 3D have been found in studies of braneworlds or holography \cite{Anber:2008qu,Astorino:2011mw,Astorino:2016xiy}. However a systematic classification \cite{Accin3D,RuthAccBH} indicated a rich panoply of solutions, with previously undiscovered branches of accelerating black holes. Three classes of solutions were found, each broadly with its own interpretation: Class I were mostly point masses, though contained a novel branch of BTZ-like black holes. Class II were accelerating analogs of the BTZ black hole, and Class III were unique asymmetric coordinatisations of AdS that led to an accelerating braneworld interpretation. A common feature of all solutions is a thread of stress energy that pulls or pushes the black hole (or point mass), thereby accelerating it. This thread is the analog of the string (or strut) in 4D, and we shall refer to it in this way although it is a domain wall in the spacetime\footnote{Note the thread has both (spatial) dimension and codimension of one, so could equally well be called a string or a wall. However, since the nomenclature {\it string / strut} is familiar from conical defects in 4D and efficiently communicates the tension of the defect, we use it when denoting a specific tension object, and reserve {\it domain wall} for the generic codimension 1 defect.}.

Acceleration therefore is obtained by an asymmetric identification of the AdS$_3$ geometry, where a domain wall is produced by identifying surfaces that have the same induced geometry, but different extrinsic curvatures. It is not immediately evident therefore how an accelerating solution in 3D can be made to rotate. In 4D, one can choose an axis of symmetry and rotate the black hole around it. In the case of acceleration, the axis is set by the cosmic string generating the acceleration -- the string itself remains invariant under the rotation, thus the rotation and acceleration act independently on the spacetime. In 3D however, the ``string'' generating the acceleration breaks the $U(1)$ symmetry of the AdS spacetime, so it is not so clear how one applies rotation as there is no invariant point around which to rotate. Further, one might be led to imagine that rotation around the black hole / point particle would cause the domain wall to also rotate, perhaps leading to superluminal rotation at sufficiently large distance from the horizon. In fact, neither of these concerns turns out to be valid: by returning to work on rotating particles and black holes in 3D we show that rotation, like acceleration (and indeed the source particles or black holes), is produced by non-trivial identifications in the spacetime. Here, we construct new families of exact solutions representing accelerating and rotating point particles, black holes, and braneworlds in 3D gravity. Our solutions have three independent parameters associated with mass, acceleration, and angular momentum, reducing to previously known 3D solutions as the rotation parameter vanishes \cite{Accin3D,RuthAccBH}.

The outline of our paper is as follows. We first review (\S \ref{sec:review}) the introduction of rotation in 3D solutions, then recap the three classes of accelerating 3D spacetimes. In \S \ref{sec:themetrics} we then construct an accelerating rotating BTZ metric using the methodology of \S \ref{sec: 3d rot} to understand the additional complexity introduced by rotation  in the geometry and motivate the introduction of a metric Ansatz that we then solve to obtain the central result of the paper: a full classification of rotating, accelerating metrics in 3D Einstein gravity. Sections \S \ref{sec:PP}-\ref{sec:C3} then study the three classes of metrics, exploring all of the possible physically distinct solutions. Finally, \S \ref{sec:wrap} concludes, and appendix~\ref{globalstructure} contains details of the transformations used to obtain the global plots of the spacetimes.

\section{Introducing rotation and acceleration}
\label{sec:review}

To frame the discussion of how to both accelerate and rotate a source in $(2+1)$-dimensions, we first 
review each of these properties in turn, showing how acceleration is manifest in both point particle and 
black hole solutions in 3D, and how (non-accelerating) point particles and black holes can be made to rotate.

Key to understanding each of these phenomena is to note the behaviour of Einstein gravity in three 
dimensions, which is topological: any solution to the vacuum Einstein field equations is locally Minkowski, 
de Sitter, or anti de Sitter (AdS), depending on whether the cosmological constant is zero, 
positive, or negative, respectively. It is this last case, AdS, that we consider here.
What distinguishes solutions to three-dimensional gravity therefore, is the global 
structure of the spacetime. One changes the global structure of a spacetime by performing 
identifications, and these identifications can be altered by changing the slicing of the spacetime.

\subsection{Rotation in three dimensions}
\label{sec: 3d rot}

We first review the construction of rotating solutions in three dimensions (3D), focussing on the
physical interpretation of coordinate choices. Consider the universal covering 
spacetime of global AdS,
\begin{equation}\label{eq: universal covering spacetime}
d s^2~=~ -\lr{\frac{R^2}{\ell^2}+1}d T^2+\frac{d R^2}{\frac{R^2}{\ell^2}+1}+R^2d\Phi^2\,,
\end{equation}
where $T\in\RR$, $R>0$, and $\Phi$ is periodic with a period to be determined. 
Static point masses in three-dimensional gravity are described by conical 
defects~\cite{Deser_massdefect,DeserConstantCurve,LoukoKepler}; 
hence, we may construct a point mass by performing the identification 
$(T,\Phi)\sim(T,\Phi+2\pi\sqrt{M_s})$, where $M_s<1$ to ensure an angular 
deficit (rather than excess). A useful set of coordinates for this identification is
\begin{equation}
T~=~\sqrt{M_s}\,t\,\quad\quad R~=~\frac{r}{\sqrt{M_s}}\,,\quad\quad \Phi~=~\sqrt{M_s}\,\phi\,,
\end{equation}
where $\phi\in(0,2\pi)$. The line element becomes
\begin{equation}
\label{eq: point mass metric}
d s^2~=~-\lr{\frac{r^2}{\ell^2}+M_s}d t^2+\frac{d r^2}{\frac{r^2}{\ell^2}+M_s}+r^2d\phi^2\,.
\end{equation}
Note, by continuing $M_s$ to negative values, \eqref{eq: point mass metric} 
describes the static BTZ black hole~\cite{BTZoriginal,BTZgeometry}, it is thus a natural
metric in which there is a continuous variation between the point particle and black hole
gravitating solutions in 3D AdS.

To construct a rotating point mass, we impose instead a helical identification~\cite{Deser_massdefect,DeserConstantCurve,LoukoKepler,SpinningParticleGeometries}  on the universal covering spacetime~\eqref{eq: universal covering spacetime},
\begin{equation}\label{eq: helical identification}
    (T,\Phi)~\sim~\lr{T-2\pi a\ell\sqrt{M_s} ,\Phi+2\pi\sqrt{M_s(1+a^2)}}\,,
\end{equation}with $a>0$ and $M_s(1+a^2)<1$. A convenient set of coordinates for this identification is
\begin{equation}
    T~=~\sqrt{M_s}\lr{\frac{t}{\sqrt{1+a^2}}-a\ell\phi}\,,\quad\quad R~=~\frac{r}{\sqrt{M_s}}\,,\quad\quad\Phi~=~\sqrt{M_s(1+a^2)}\,\phi\,,
\end{equation}where $\phi\in(0,2\pi)$. The associated line element is
\be\label{eq: rotating point particle ads}
\beal
d s^2~=~ &-\lr{\frac{r^2}{\ell^2}+M_s}\frac{d t^2}{1+a^2}
+\frac{d r^2}{\frac{r^2}{\ell^2}+M_s}+\frac{2a\ell}{\sqrt{1+a^2}}
\lr{\frac{r^2}{\ell^2}+M_s}d td\phi\\
& +\lr{r^2-a^2\ell^2M_s}d\phi^2\,.  
\eeal\ee
\begin{figure}
\centering
    \begin{subfigure}[b]{0.325\textwidth}
    \centering
    \includegraphics[width=\textwidth]{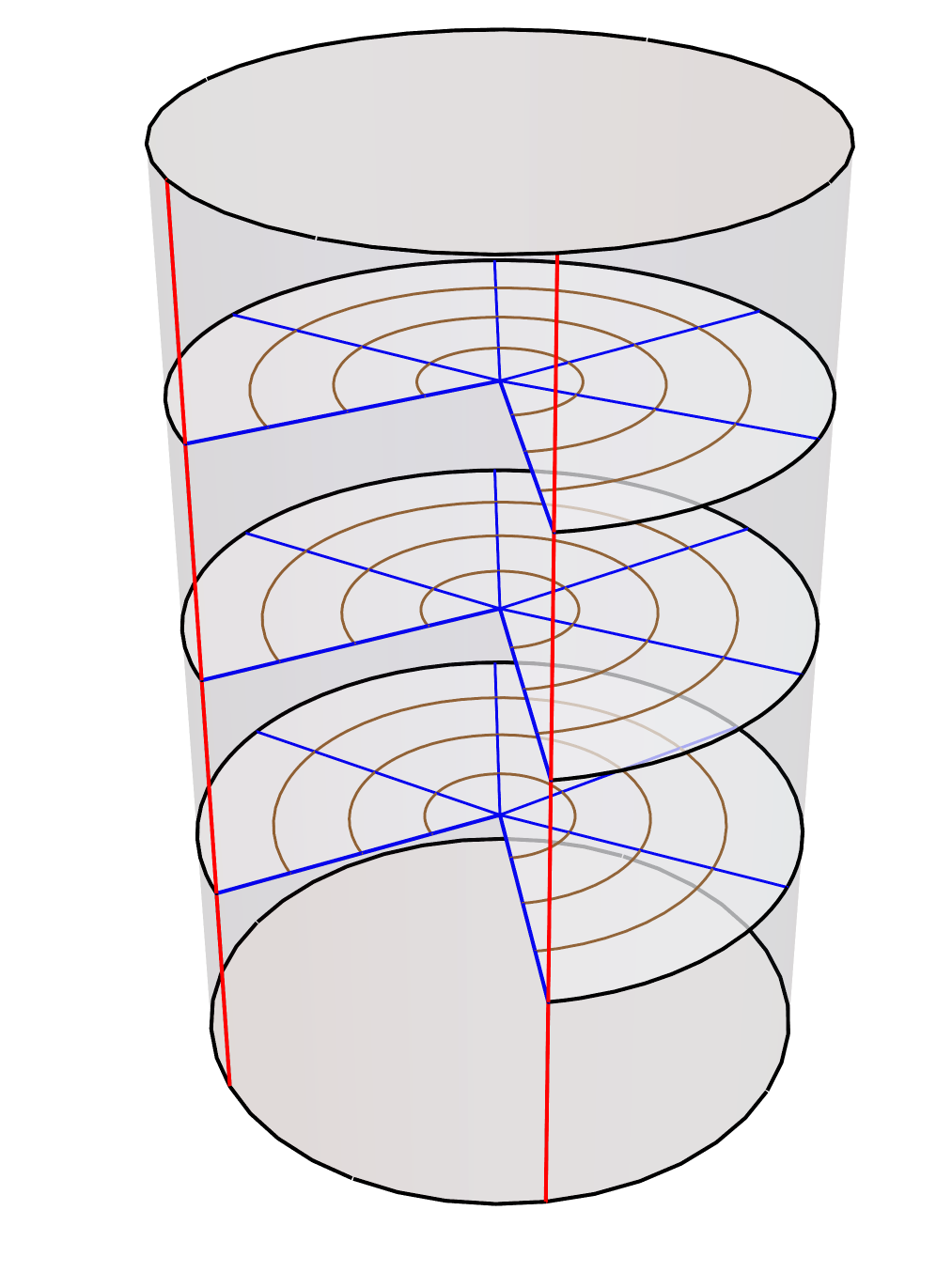}
    \caption{\label{fig:PtPcl1}}
    \end{subfigure}
    \begin{subfigure}[b]{0.323\textwidth}
    \centering
    \includegraphics[width=\textwidth]{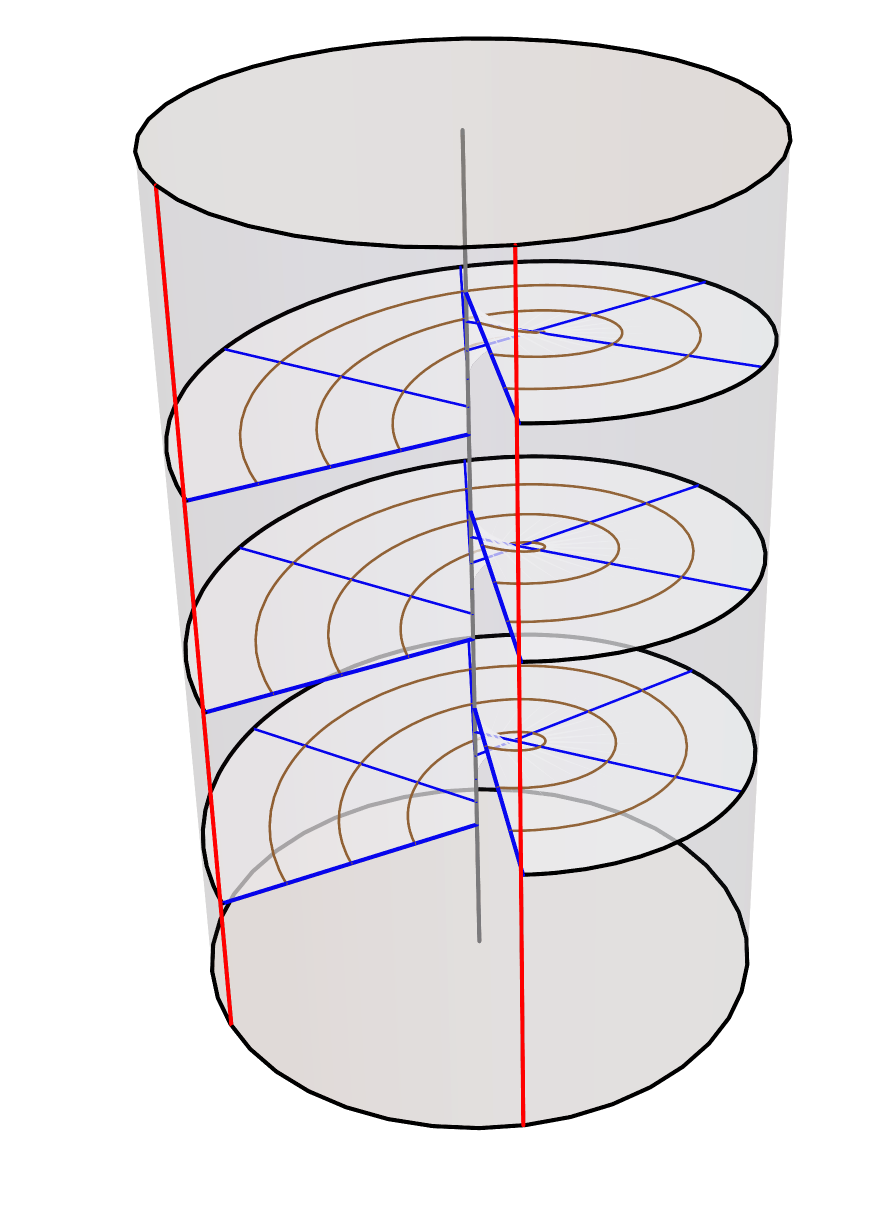}
    \caption{\label{fig:PtPcl2}}
    \end{subfigure}
    \begin{subfigure}[b]{0.32\textwidth}
    \centering
    \includegraphics[width=\textwidth]{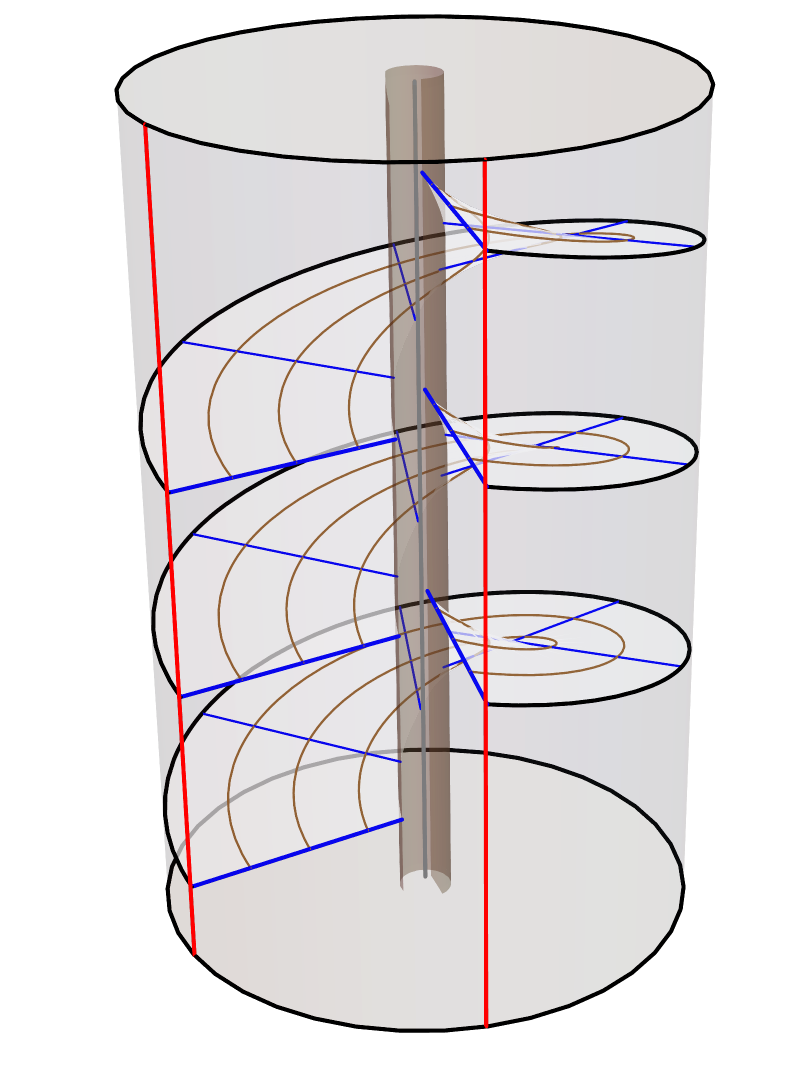}
    \caption{\label{fig:PtPcl3}}
    \end{subfigure}
\caption{Point particles in the AdS cylinder. In each  figure, three constant-$t$ slices 
are shown. Brown lines are constant $r$, blue lines constant $\phi$, red lines 
where removed conical wedge meets the boundary, and black lines where the 
constant-$t$ surface meets the boundary. 
(\ref{fig:PtPcl1}) Static point particle~\eqref{eq: point mass metric}. 
(\ref{fig:PtPcl2}) Rotating point particle~\eqref{eq: rotating point particle ads}; 
the helical structure of the identification~\eqref{eq: helical identification} 
can be clearly seen. 
(\ref{fig:PtPcl3}) Rotating point particle~\eqref{eq: rotating point particle ads} 
with the region of closed timelike curves $r\leq r_{\chi}$ excised from 
the centre.\label{fig:ParticleTriptych}}
\end{figure}

The causal structure of the rotating spacetime~\eqref{eq: rotating point 
particle ads} is remarkably different from its static counterpart~\eqref{eq: point mass metric}. 
Denoting by $g_{\mu\nu}$ the metric components of~\eqref{eq: rotating point particle ads}, 
we see that the component $g_{\phi\phi}$ changes sign at
\begin{equation}
    r_{\chi}~=~a\ell\sqrt{M_s}\,.
\end{equation}
In particular, $g_{\phi\phi}$ is negative for $r<r_{\chi}$, indicating the presence of 
closed causal curves; within this region, the global identification becomes timelike. 
This identifies $r_{\chi}$ as a {\it chronology} horizon. Physically reasonable spacetimes 
do not contain closed causal curves~\cite{HawkingChronologyProtection} and so we 
excise this region from our spacetime
\cite{BTZoriginal,BTZgeometry,SpinningParticleGeometries,MassEntropydeSitter,KdS3D}
by introducing the radial coordinate $\rho^2=g_{\phi\phi}$ and restrict our 
attention to the region $\rho^2\geq0$. In coordinates $(t,\rho,\phi)$, the line 
element~\eqref{eq: rotating point particle ads} becomes
\begin{subequations}\label{eq: rotating point particle BTZ}
\begin{align}
d s^2&~=~-f(\rho)d t^2+\frac{d\rho^2}{f(\rho)}+\rho^2\lr{d\phi+\frac{J}{2}\lr{\frac{1}{\rho^2}
+\frac{1}{M_s\ell^2(1+a^2)}}d t}^2\,,\\
f(\rho)&~=~\frac{\rho^2}{\ell^2}+M+\frac{J^2}{4\rho^2}\,,
\end{align}
\end{subequations}
where $M=(1+2a^2)M_s$ and $J=2a\ell M_s\sqrt{1+a^2}$ are the rotating 
mass and angular momentum respectively. We note that again, by 
continuing $M_s$ to negative values, \eqref{eq: rotating point particle BTZ} 
becomes the line element of the rotating BTZ black hole of mass $|M|$. 

The spatial section of the point particle metrics can be compactified and 
mapped into the Poincaré disk~\cite{Accin3D}; this compactification alongside 
the time coordinate allows one to map these point particle geometries into a 
cylinder. In Fig.~\ref{fig:ParticleTriptych}, we plot three constant-$t$ slices 
each for the static point particle~(\ref{fig:PtPcl1}), the rotating point particle
(\ref{fig:PtPcl2}), and the rotating point particle with the region of closed 
timelike curves excised~(\ref{fig:PtPcl3}).

Finally, we note that the line element~\eqref{eq: rotating point particle BTZ} is in a 
frame such that spatial infinity is rotating,
\begin{equation}\label{eq: horizon rotation}
    \lim_{\rho\to\infty}-\frac{g_{t\phi}}{g_{\phi\phi}}~=~-\frac{a}{\ell\sqrt{1+a^2}}\,.
\end{equation}
We refer to this as the corotating frame since, in the case of the rotating BTZ 
black hole ($M_s<0$), the outer event horizon is static in this frame. 

\subsection{Acceleration}

We now review the static, or non-rotating, C-metric in 3D \cite{Accin3D,Anber:2008qu,Astorino:2011mw,Xu:2011vp,Astorino:2016xiy}.
Whilst it is common to refer to ``the'' three-dimensional C-metric, the reality is a trio of 
families of solutions, first fully characterised in~\cite{Accin3D}, parametrised by an 
acceleration parameter $A$, dubbed Class I, Class II, and Class III according to the
functional form of the solution. The line element  can be given in either 
canonical $\{\tau,x,y\}$ \cite{Accin3D}, or polar coordinates $\{t,r,\phi\}$ \cite{BunneyNutshell},
by
\begin{equation}
\beal
\label{eq: C-metric general non rotate}
d s^2~&=~\frac{1}{A^2(x-y)^2} \left ( - P(y) d \tau^2 + \frac{d y^2}{P(y)} + \frac{d x^2}{Q^2 (x)}
\right ) \\
&=~\frac{1}{\lr{1+\mc{A}r\Xi(m\phi)}^2}\lr{-f(r)d t^2+\frac{d r^2}{f(r)}+r^2d\phi^2}
\eeal
\end{equation}
The various metric functions are given in Table~\ref{table:ClassParameters} 
with the transformation between the coordinate systems being 
$\bar{t} = m^2 \mc{A} t$, $y = -1/\mc{A}r$, and 
$x = \pm \Xi(m\phi)$ with $A = m\mc{A}$, and the two branches of the 
angular identification come into play during the construction of 
accelerating solutions where the coordinate $\phi$ spans the
part of the spacetime that is retained.

Canonical coordinates are most useful for classification of the solutions and 
particularly in understanding their global structure, whereas the polar system is 
perhaps more intuitive, and displays the continuity of point particle and black hole solutions via
the introduction of what, at this point, is a gauge parameter $m$. Each system will be seen to play
a role in the construction and understanding of the rotating accelerating solutions.
For now, we note that the AdS boundary that would be at $r\to\infty$ for $\mc{A}=0$
is replaced by $x=y$, or $\Omega = 0$ where $\Omega(r,\phi)=1+\mc{A}r\Xi(m\phi)$. 
The anti-de Sitter length scale is given by $\ell$.
\begin{table}[t!]
\centering
\begin{tabular}{|c|c|c||c|c|c|} 
\hline
Class & $f(r)$ & $\Xi(\psi)$ & $P(y)$ & $Q^2(x)$ & $x$ range\\ [0.5ex] 
\hline
I & $\frac{r^2}{\ell^2}+m^2(1-\mc{A}^2r^2)$ & $\cos(\psi)$ 
& $\frac{1}{A^2\ell^2}+(y^2-1)$ & $1-x^2$ & $|x|\leq1$ \\ [0.5ex]
II & $\frac{r^2}{\ell^2}-m^2(1-\mc{A}^2r^2)$ & $\cosh(\psi)$ 
& $\frac{1}{A^2\ell^2}+(1-y^2)$ & $x^2-1$ & $|x|\geq1$ \\[0.5ex]
III & $\frac{r^2}{\ell^2}-m^2(1+\mc{A}^2r^2)$ & $\sinh(\psi)$ 
& $\frac{1}{A^2\ell^2}-(1+y^2)$ & $1+x^2$  & $\mathbb{R}$ \\ [0.5ex] 
\hline
\end{tabular}
\caption{The three classes of static C-metric.}
\label{table:ClassParameters}
\end{table}

The construction of these C-metric solutions is outlined in~\cite{Accin3D,RuthAccBH}. 
The various classes of metric represent different gauge transformations of AdS$_3$ 
parametrized by $A$, that ``skew'' the coordinates in a particular fashion. Classes I and II
have a reflection symmetry, $\phi \leftrightarrow -\phi$ (realised in canonical coordinates
by a patching - see below) and are then cut along mirror image surfaces and identified. 
Class III on the other hand has no such symmetry, and both
polar and canonical coordinates cover the Poincar\'e patch in an asymmetric fashion.
This leads to a qualitatively different accelerating spacetime as described in \S \ref{sec:C3}.

In canonical coordinates, two mirror image copies of the Class I or II spacetime are glued
along $|x|=1$ where the extrinsic curvature of that boundary vanishes; the resulting spacetime
now fully covers AdS$_3$ (or it's Poincar\'e patch).
Two second boundaries at $x=x_0$ are then chosen, one on each side, 
that now have nonzero but equal and opposite extrinsic curvatures; these are then glued 
together resulting in a codimension one defect with either positive (string) 
or negative (strut) tension that is
derived via the Israel junction conditions \cite{Israel_thin_shell}.

A quick way to note whether the acceleration is driven by a string or strut is to 
note how the warp factor ($\Omega^{-1}$) behaves away from the wall. 
Positive tension is indicated by a local peak in the
warp factor and negative tension by a local trough. 
Strings are therefore characterised by $x>x_0$
in the geometry and struts by $x<x_0$.
In polar coordinates there is no need to identify two mirror images as the $\phi-$coordinate
naturally spans the full range of the spacetime, which is now more simply found by identifying 
$\phi=\pi$ with $\phi = -\pi$; this then reveals the interpretation
of the $m$ parameter via $x_0=\Xi[m\pi]$. This identification comes with 
a specific sign for the tension; the opposite tension 
can be obtained by changing the sign in the relation between $x$ 
and $\phi$, or by the simple expedient of swapping the sign 
of $\mc{A}$ in the polar metric. 
For Class III there are no angular surfaces with vanishing extrinsic curvature -- both
coordinate systems fully cover the Poincar\'e patch, therefore we must take \emph{two} 
copies of AdS$_3$ in Class III coordinates, cut along a constant $x$ or $\phi$ surface in each,
retain either $x<x_0$ or $x>x_0$ and identify the two sides -- a construction reminiscent of
the Randall-Sundrum braneworld \cite{Randall:1999ee,Randall:1999vf}.

Class I metrics describe, in the main, a point 
particle in AdS accelerated by the domain wall. 
For point particles, the $y-$coordinate is taken to be 
negative and bounded above by either $x$, or the value at which $P(y)=0$. 
The `origin', around which there is a conical deficit
defining the point particle, is where $g_{xx} \to 0$, 
i.e.\ $y\to-\infty$. The domain wall at $x_0$ then defines
both the conical defect angle and the accelerating string/strut tension.
This interpretation is made manifest in polar coordinates by noting 
that the periodicity of the spacetime is $2\pi m$. 
The parameter $m$, taking values $0\leq m<1$, determines the size of 
the conical defect at the origin. The ``mass'' of this defect is 
given by the conical deficit $\delta = 2\pi - 2\pi m$~\cite{Deser_massdefect}
\begin{equation}
m_c~=~\frac14(1-m)\,,
\end{equation} 
whilst the tension $\sigma$ of the string is
\begin{equation}
\sigma~=~\frac{m\mc{A}}{4\pi}\sin(m\pi)\,.
\end{equation}
A point defect accelerated or pushed by a strut 
is obtained by identifying instead $x=-\cos(m\phi)$, in which case 
$\sigma$ flips sign.

Within the Class I metrics, it is also possible to take $y>0$ if we choose
$x_0\in(y_h,1)$, where $P(y_h)=0$. This gives rise to an accelerating 
black hole as it is no longer possible for $g_{xx}=0$;  thus there is
no point particle, but there is a compact horizon.

The Class II family describes a black hole in AdS, 
accelerated by a string or strut. 
The black-hole horizon is located at
\begin{equation}
r_h~=~\frac{m\ell}{1+m^2\mc{A}^2\ell^2}\,.
\end{equation}
The mass of the black hole is a non-trivial function of $m\geq 0$ and $\mc{A}$~\cite{RuthAccBH} 
that vanishes in the limit $m\to0$ and tends to $m^2$ as $\mc{A}$ tends to zero. 
The tension of the strut (obtained from the natural choice of $\phi(x)$) is
\begin{equation}
\sigma~=~-\frac{m\mc{A}}{4\pi}\sinh(m\pi)\,.
\end{equation}
To obtain a black hole accelerated by a string, one simply
identifies $x = - \cosh (m\phi)$, in which case one would be considering negative
$x_0$ in the canonical system, or flips the sign of $\mc{A}$ if remaining within the 
polar description of the spacetime.

Turning to the Class III solutions that were briefly mentioned in \cite{Accin3D}, the
canonical coordinates easily show that $x$ is limited by $x>y>-y_h$, hence we can either 
include the boundary by taking $x_0<y_h$, or exclude it by choosing $x_0>y_h$. 
Within the former choice however there are another 2 options: one can keep the 
spacetime with $x<x_0$ or $x>x_0$. These correspond to a strut or string solution
respectively. Indeed, one can construct a geometry with both a string and strut
by choosing two surfaces, $x_0$ and $x_1$, and gluing two copies of the segment
$x_0<x < x_1$ together in a construction reminiscent of the original (RS1) Randall-Sundrum 
braneworld scenario \cite{Randall:1999ee}.

\section{The rotating C-metric}
\label{sec:themetrics}

Having reviewed both the rotation and acceleration in isolation, we now combine the 
conceptual approaches in this section to derive the full classes of rotating acceleration 
solutions in three dimensions: the rotating C-metrics.

\subsection{The rotating BTZ solution}
Before classifying the rotating C-metric in full generality, we develop intuition by applying the procedure of \S\ref{sec: 3d rot} to the accelerating BTZ Class II C-metrics:
\be
\label{eq: static C metric II}
d s^2 ~=~\frac{1}{(1+\mc{A}R\cosh(m\Phi))^2}\lrb{-f(R)d T^2+\frac{d R^2}{f(R)}+R^2d\Phi^2}\,,
\ee
where $f(R)= R^2/\ell_m^2-m^2$, and we have introduced an effective AdS length scale 
$\ell_m^2=\ell^2/(1+\mc{A}^2\ell^2m^2)$ for notational convenience.
Cast in this form, the static metric appears conformal to the BTZ black 
hole of mass $m^2$ and cosmological constant $-1/\ell_m^2$~\cite{BTZoriginal,BTZgeometry}; 
however crucially at this stage of the discussion we have not compactified $\Phi$.

We first perform a boost in $(T,\Phi)$ to new coordinates $(t,\theta)$ via
\begin{equation}
    \begin{pmatrix}
        T\\\ell_m\Phi
    \end{pmatrix}~=~\begin{pmatrix}
        b&-a\\-a&b
    \end{pmatrix}\begin{pmatrix}
        t\\\ell_m\theta
    \end{pmatrix}\,,
\end{equation}
where $a\in\RR$ and $b\geq1$ such that $b^2-a^2=1$, and the sign of $b$ is chosen to preserve time orientation. Under the action of this boost, the metric~\eqref{eq: static C metric II} may be written as
\be
\beal
ds^2~ &=~\frac{1}{\lr{1+\mc{A}R\cosh\lr{mb\theta-m\frac{a}{\ell_m}t}}^2}
\Bigg[-\bigg(\frac{R^2}{\ell_m^2}-(1+a^2)m^2\bigg)dt^2\\
& \qquad\qquad\qquad\qquad
+\frac{dR^2}{f(R)}
+(R^2+a^2\ell_m^2m^2)d\theta^2-2ab\ell_m m^2d\theta dt\Bigg]\,.
\eeal
\ee
Introducing the radial coordinate $\rho^2=R^2+a^2\ell_m^2m^2$, we obtain the metric
\begin{align}\nonumber
    ds^2~=~\frac{1}{\lr{1+\mc{A}\sqrt{\rho^2-\rho_-^2}\cosh\lr{mb\theta-m\frac{a}{\ell_m}t}}^2}&\Bigg[-g(\rho)dt^2\\
    +\frac{d\rho^2}{g(\rho)}&+\rho^2\lr{d\theta-\frac{J}{2\rho^2}dt}^2\Bigg]\,,\label{eq:Rot BTZ conf}
\end{align}
where
\begin{align}
M=m^2(1+2a^2)\,,\qquad J=2a\ell_m m^2\sqrt{1+a^2}\,,\qquad g(\rho)=\frac{\rho^2}{\ell_m}-M+\frac{J^2}{4\rho^2}\,,
\end{align}and $\rho_-=a\ell_mm$, $\rho_+=\ell_mm\sqrt{1+a^2}$ are the positive zeros of $g(\rho)$. The line element~\eqref{eq:Rot BTZ conf} is conformal to that of a rotating BTZ black hole~\cite{BTZoriginal,BTZgeometry,CarlipBTZ} of mass $M$, angular momentum $J$, and cosmological constant $-1/\ell_m^2$. Just as in the case of the BTZ black hole, we may express $M$ and $J$ purely in terms of $\rho_\pm$ and the effective AdS length scale $\ell_m$,
\begin{equation}
    M~=~\frac{\rho_+^2+\rho_-^2}{\ell_m^2}\,,\qquad J~=~2\frac{\rho_+\rho_-}{\ell_m}\,.
\end{equation}

To address the time dependence of the conformal factor, we introduce the corotating coordinate
\begin{equation}
    \theta~=~\phi+\frac{a}{\ell_m\sqrt{1+a^2}}t\,,
\end{equation}arriving at the metric
\be
\label{eq:rotating Class II}
d s^2=\frac{1}{\Omega_a^2} \left [ -g(\rho)d t^2+\frac{d \rho^2}{g(\rho)}
+\rho^2\lr{d\phi+\frac{J}{2}\lr{\frac{1}{\rho_+^2}-\frac{1}{\rho^2}}\!d t}^2\right ]\,,
\ee
with
\be
\qquad\Omega_a = \left [1+\mc{A}\sqrt{\rho^2-\rho_-^2}\cosh(m\sqrt{1+a^2}\,\phi) \right]\,.
\ee
To complete the construction, we identify $(t,\phi)\sim(t,\phi+2\pi)$.

The global structure of this spacetime is distinct from the non-rotating BTZ Class II C-metric~\eqref{eq: static C metric II}, which we may see as follows. The identification imposed to construct the accelerating BTZ spacetime~\eqref{eq: static C metric II} is $(T,\Phi)\sim(T,\Phi+2\pi)$. If we work back through the coordinate transformations in this section, the identification $(t,\phi)\sim(t,\phi+2\pi)$ may be written in the original coordinates as the helical identification
\begin{equation}
    (T,\Phi)~\sim~(T-2\pi a\ell_m,\Phi+2\pi\sqrt{1+a^2})\,,
\end{equation}with the two identifications agreeing only in the non-rotating limit as $a$ tends to zero.

Given the rotating C-metric~\eqref{eq:rotating Class II}, it is clear that one could apply the same techniques to 
generalise the Class I and Class III solutions. However, to ensure that we cover the full solution space, 
we go back to the beginning and solve Einstein's field equations.

\subsection{Rotating C-metric \textit{ab initio}}

We return to the methodology of \cite{Accin3D} and consider an Ansatz in canonical coordinates $(\tau,x,y)$. 
Inspecting the rotating metrics~\eqref{eq: rotating point particle BTZ} and~\eqref{eq:rotating Class II}, 
we see that to include rotation, any Ansatz should include $t$--$x$ cross terms in the metric. 
Therefore, motivated by the previous discussion, we consider the following metric ansatz:
\begin{equation}\label{eq:canon rot form}
d s^2 ~=~ \frac{1}{{A}^2(x-y)^2} \left [ -\frac{P(y)}{H(y)} d \tau^2 + \frac{d y^2}{P(y)}
+ H(y) \left ( \frac{d x}{Q(x)} + j \frac{P(y)}{H(y)} d \tau \right)^2 \right]
\end{equation}

Using Einstein's field equations, coordinate rescaling, and re-zeroing as in~\cite{Accin3D}, 
the metric functions can be found to take the following form
\be
\beal
Q^2(x) &~=~ \epsilon_2 x^2 + \epsilon_0\,, \\
H(y) &~=~ h_0 +\epsilon_2j^2  y^2\,, \\
P(y) &~=~ \frac{1}{ {A}^2\ell^2}  - \frac{({A}^2\ell^2 - j^2)}{{A}^2\ell^2(h_0-j^2 \epsilon_0)} Q^2(y)\,,
\eeal
\ee
where the choices of $\epsilon_{0,2} = \pm 1$ determine the Class of the 
metric.

It is clear that the parameter $j$ represents the rotational
degree of freedom in the metric, but it would appear that $h_0$ is an
additional degree of freedom. However, there still remains
a further rescaling degree of freedom in this solution
\be
j \to \lambda j \;,\;\; h_0 \to \lambda^2 h_0 \;,\;\;  A \to \lambda A \;
\Rightarrow  \;\; H \to \lambda^2 H \;, \;\; P \to P/\lambda^2 \;, \;\; \tau \to \lambda^3 \tau\,.
\ee
This freedom allows us to choose a specific value of $\lambda$ to set
\be
\label{eq:canonmetfns}
\beal
Q^2(x) &~=~ \epsilon_2 x^2 + \epsilon_0\,, \\
H(y) &~=~ 1 - \frac{j^2}{A^2\ell^2} + j^2 Q^2(y)\,, \\
P(y) &~=~ \frac{1}{ {A}^2\ell^2}  - Q^2(y)\,. 
\eeal\ee
This final form of the metric functions now has one additional degree of freedom, 
represented by the parameter $j$ that introduces the $t$--$x$ cross term, and also
makes the function $H$ nontrivial. 
The three classes of solution are summarized in Table~\ref{tab:classes}. 

\begin{table}[t]
\caption{The three classes of rotating C-metrics.}
\centering
\begin{tabular}{ | c || c | c | c | c | c | c | c |}
\hline 
Class   & $\epsilon_0$ & $\epsilon_2$ & $Q^2(x)$ & $H(y)$  & $P(y)$    &  $x$ range \\
\hline\hline 
~  & & & & & &\\
I & $+$       & $-$  & $1-x^2$ & $1 - \frac{j^2}{A^2\ell^2} + j^2 (1-y^2)$  
& $\frac{1}{A^2\ell^2}- (1-y^2)$ & $|x|<1$   \\
~  & & & & & &\\
II & $-$       & $+$ &  $x^2-1$  & $1 - \frac{j^2}{A^2\ell^2} + j^2 (y^2-1)$
& $\frac{1}{A^2\ell^2}- (y^2-1)$ & $|x|>1$   \\
~  & & & & & &\\
III & $+$   & $+$  & $x^2+1$ & $1 - \frac{j^2}{A^2\ell^2} + j^2 (1+y^2)$ 
& $\frac{1}{A^2\ell^2}- (y^2+1)$ & $x \in \mathbb{R}$   \\
~  & & & & & &\\
\hline
\end{tabular}
\label{tab:classes}
\end{table}

Having derived this form of the metric however, the observant reader will notice that a rearrangement of
the metric \eqref{eq:canon rot form} is
\be
ds^2 = \frac{1}{A^2 (x-y)^2} \left [ - P(y) \left ( d\tau - j \frac{dx}{Q(x)} \right )^2 
+ \frac{dy^2}{P(y)} + \frac{dx^2}{Q^2(x)} \right]
\label{transform-to-t}
\ee
in other words, a boost, or redefinition of the time coordinate in the original canonical metrics has
taken place in much the same way as discussed in  \S\ref{sec: 3d rot}. 
It is tempting to set $dt = d\tau - j dx/Q$. However while this might return us to the canonical
C-coordinates in terms of the full AdS spacetime,
once one makes cuts or identifications (as we discuss in the next section) this identification 
{\it at constant $\tau$} gives a new, rotating, spacetime.

\subsection{Acceleration and defect tension}

Thus far, the various Classes of metrics are simply a rewriting of AdS$_3$ in a skewed system.
In order to construct a physically new spacetime, we must make identifications and excise parts of
the global anti-de Sitter geometry. This is done by cutting along a constant $x-$line and identifying
with a mirror image as described in \cite{Accin3D,RuthAccBH}. The resulting string or strut
provides the ``force'' for acceleration. 
The metric on and parallel to the $x_0-$surface only depends on $x$ via the conformal factor; 
intuitively, positive (negative) energy/tension corresponds to a local peak (dip) 
of this warp factor. In each case, the normal points {\it out} of the spacetime on the {\it minus} 
side in the Israel equations, and {\it into} the spacetime on the {\it plus} side.

Given the canonical rotating form~\eqref{eq:canon rot form}, the normal form is
\be
\vb{n}~=~ \frac{\epsilon_nd x}{{A} (x-y) Q}\,,
\ee
where $\epsilon_n=\pm1$ indicates whether the normal points towards increasing or 
decreasing $x$, and we have used the relation $H(y)=1-j^2P(y)$ 
(cf.~\eqref{eq:canonmetfns}). We project to intrinsic coordinates $(T,Y)$ on the $x=x_0$ surface,
\be
X^\mu ~=~ (T,x_0,Y) \quad \Rightarrow \qquad 
X^\mu_{,A} ~=~ \left \{ \begin{matrix}
1 & 0 & 0 \\
0 & 0 & 1
\end{matrix} \right \}\,,
\ee
leading to the intrinsic metric
\be
\gamma_{AB} ~=~
 \frac{1}{{A}^2(x-y)^2} \left ( \begin{matrix}
- P& 0 \\
0 & \frac1P
\end{matrix} \right )\,.
\ee
The extrinsic curvature is then
\be 
K_{AB} ~=~ X^a_{,A} X^b_{,B} \nabla_a n_b ~=~ \epsilon_n AQ(x_0) \gamma_{AB}\,,
\ee from which form we may read off the tension as
\begin{equation}
\label{eq: sigmadef}
\sigma~=~\epsilon_n \frac{A}{4\pi}Q(x_0)\,.
\end{equation}
Interestingly, this result is the same as in~\cite{Accin3D}, and indeed the expression 
in \eqref{transform-to-t} indicates that from a global perspective, the defect at $x=x_0$ is
the same defect whether or not there is rotation. However, recall we are identifying
at constant $\tau$ -- this is crucial as constant $\tau$ is actually at different global time
as we will see once we delve into the structure of these metrics more closely.

\section{Accelerating and rotating point particles}
\label{sec:PP}

Point sources in 3D gravity are represented by local conical deficits, 
which can be viewed as an identification of two surfaces, each at constant
angle, emerging from the point itself. An accelerating point particle happens
when these constant angle surfaces no longer have zero extrinsic curvature; a
rotating point particle is when there is an angularly dependent shift applied 
to the time coordinate prior to identification. 
The accelerating and rotating point particle emerges when both procedures are 
applied simultaneously as described in the previous section to the Class I metrics.

From Table~\ref{tab:classes}, we see that the function $P(y)$ may exhibit zero, 
one, or two distinct real roots, depending on whether $A\ell<1$, $A\ell=1$, 
or $A\ell>1$ respectively. Following the conventions for the different 
phases of acceleration of the static C-metric in~\cite{Accin3D}, we refer 
to these three cases as slow ($A\ell<1$), saturated ($A\ell=1$), and rapid ($A\ell>1$).

\subsection{Slow acceleration}
\label{sec: slow pp}

\begin{figure}
\centering
    \begin{subfigure}[b]{0.49\textwidth}
    \centering
    \includegraphics[width=\textwidth]{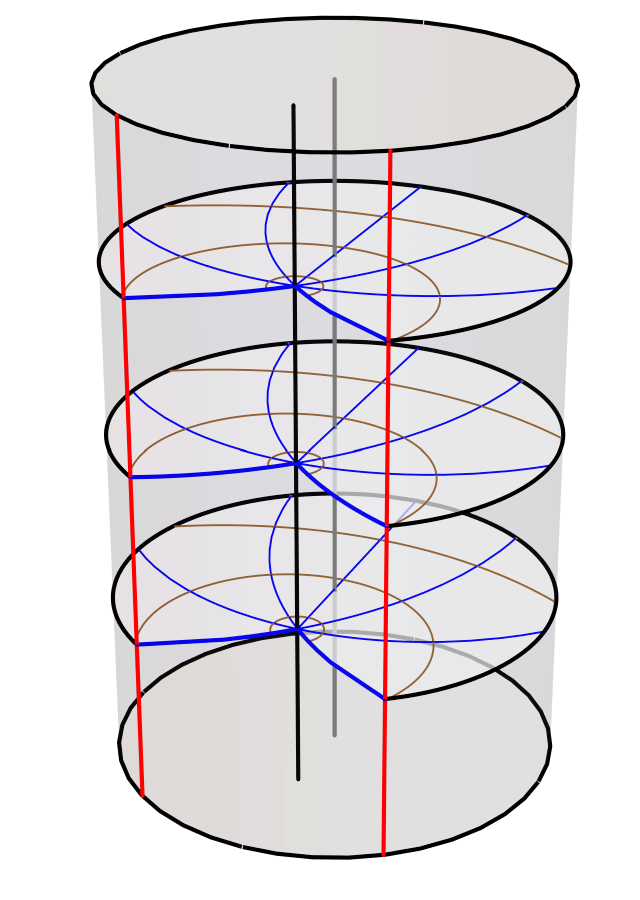}
    \caption{\label{fig:1Slow}}
    \end{subfigure}
    \begin{subfigure}[b]{0.49\textwidth}
    \centering
    \includegraphics[width=\textwidth]{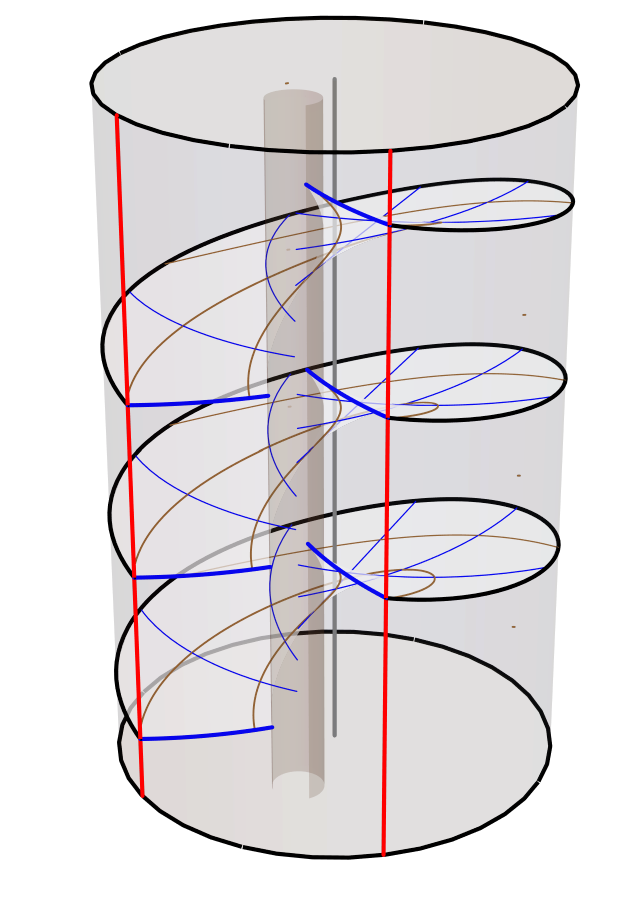}
    \caption{\label{fig:1Slow2}}
    \end{subfigure}
\caption{Slowly accelerating point particles in AdS mapped onto global AdS3. 
In each  figure, three constant-$t$ slices are shown. Brown lines are constant $r$ or $y$, 
blue lines constant $x$ or $\phi$, red lines where the removed conical wedge meets the 
boundary, and black lines where the constant-$t$ surface meets the boundary. 
(\ref{fig:1Slow}) Accelerating point particle ($J=0$)~\eqref{eq:Class I rot slow}. 
(\ref{fig:1Slow2}) Accelerating and rotating point particle~\eqref{eq:Class I rot slow} 
with brown central region denoting the excised region of closed timelike curves.
\label{fig:Class1Slow}}
\end{figure}

For the slowly accelerating particle, $P(y) = y^2 -1 + 1/A^2 \ell^2 >0$ for all $A\ell<1$.
The canonical construction takes two copies of the spacetime, gluing along $x=1$ and
$x=x_0$ ($x\in[x_0,1]$) as described in the previous section. The primary patch and its mirror image
cover the full spacetime, with the shifted identification introducing rotation as indicated in
figure \ref{fig:Class1Slow}. The choice of $x_0$ is readily seen to determine both the
mass of the conical point deficit, as well as the tension of the string providing the 
force for acceleration. 
 The chronology limiting surface, where $\phi$ becomes null, is identified as $H\to0$. To better represent
this spacetime, we 
introduce the following polar coordinates
\begin{equation}
\label{eq: C I slow polar}
\tau~=~\frac{m^2\mc{A}t}{\sqrt{1+a^2}}\,,\quad 
x~=~\cos(m\sqrt{1+a^2}\,\phi)\,,\quad y~=~-\frac{1}{\mc{A}\sqrt{\rho^2+a^2\ell_m^2m^2}}\,,
\end{equation}
where $t\in\RR$, $\rho>0$, $\phi\in(-\pi,\pi)$;  we have set $A=\mc{A}m$, 
$\ell_m^2=\ell^2/(1-\mc{A}^2\ell^2m^2)$, and $j^2=a^2\mc{A}^2\ell_m^2m^2/(1+a^2)$, 
and we require $m\sqrt{1+a^2}<1$. (Note, this choice of $x$ corresponds to 
a positive-tension solution.)
The slow acceleration phase can now be seen to be a one-parameter extension of the 
metric \eqref{eq: rotating point particle BTZ} of the
rotating point 
particle in AdS.  

The line element of the Class I rotating C-metric then reads
\begin{subequations}\label{eq: I rot slow full}
\begin{align}\label{eq:Class I rot slow}
d s^2&~=~\frac{1}{\Omega^2(\rho,\phi)}\lrb{-g(\rho)d t^2+\frac{d\rho^2}{g(\rho)}
+\rho^2\lr{d\phi+\frac{J}{2}\lr{\frac{1}{\rho^2}+\frac{1}{m^2\ell_m^2(1+a^2)}}d t}^2}\,,\\
\label{eq: C I slow Omega}
\Omega(\rho,\phi)&~=~1+\mc{A}\sqrt{\rho^2+a^2\ell_m^2m^2}\cos(m\sqrt{1+a^2}\,\phi)\,,\\
g(\rho)&~=~\frac{\rho^2}{\ell_m^2}+M+\frac{J^2}{4\rho^2}\,,
\end{align}
\end{subequations}
where $M=m^2(1+2a^2)$ and $J=2a\ell_mm^2\sqrt{1+a^2}$. Indeed, in the limit 
$\mc{A}\to0$, the line element~\eqref{eq:Class I rot slow} tends to that of a rotating 
point particle in AdS~\eqref{eq: rotating point particle BTZ}. 
The tension of the domain wall is given by
\begin{equation}
    \sigma~=~\frac{\mc{A}m}{4\pi}\sin(m\pi\sqrt{1+a^2})\,.
\end{equation}

In canonical coordinates ($\tau, x, y$), the boundary of global AdS is
located at $x=y$, and for the slowly accelerating rotating particle there is always
a segment of the boundary included in the spacetime either for $x>x_0$ (positive
tension) or $x<x_0$ (negative tension). However, in polar coordinates, 
\eqref{eq: C I slow polar}, whether the chart includes the boundary depends
on the relative magnitudes of the parameters ($a,\mc{A},\ell,m$). 
For $m\sqrt{1+a^2}<1/2$, $x_0 = \cos(m \sqrt{1+a^2} \pi)>0$, and since
$y\leq0$ for $\rho \in (0,\infty)$, the chart does not extend to the boundary. 
For $1/2<m\sqrt{1+a^2}<1$, the conformal boundary is located at
\begin{equation}
    \rho^2_{\mrm{C}}~=~\frac{1}{\mc{A}^2\cos^2(m\sqrt{1+a^2}\,\phi)}-a^2\ell_m^2m^2\,,
\end{equation}
for $m\sqrt{1+a^2}|\phi|\in\lr{\frac{\pi}{2},\arccos(-\tfrac{1}{a\mc{A}\ell_mm})}$,
and the polar coordinates~\eqref{eq: C I slow polar} do not cover the full 
spacetime. To cover the full spacetime, one would have to introduce a second 
chart covering $y>0$. In the polar coordinates~\eqref{eq: I rot slow full}, the 
effect of this is to flip the sign in front of the square root in~\eqref{eq: C I slow Omega}.

\begin{figure}
\centering
    \begin{subfigure}[b]{0.49\textwidth}
    \centering
    \includegraphics[width=\textwidth]{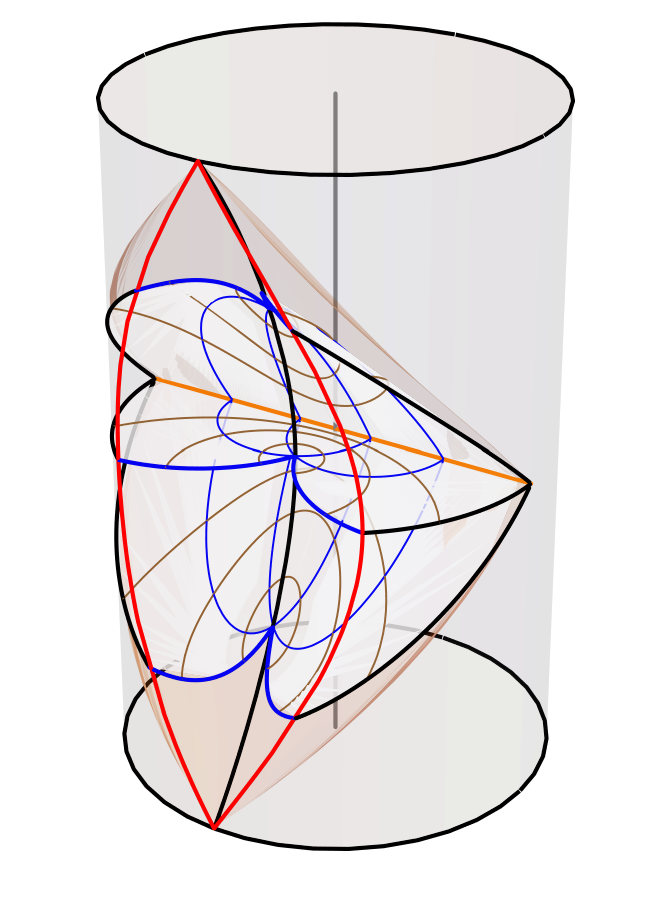}
    \caption{\label{fig:FastLight1}}
    \end{subfigure}
    \begin{subfigure}[b]{0.49\textwidth}
    \centering
    \includegraphics[width=\textwidth]{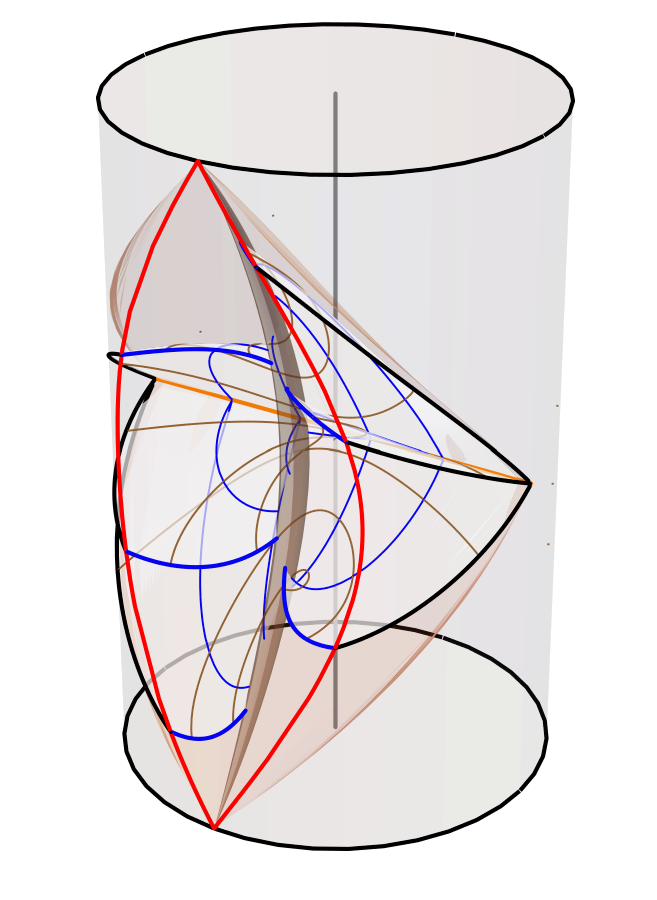}
    \caption{\label{fig:FastLight2}}
    \end{subfigure}
\caption{Light, rapidly accelerating point particles in AdS mapped onto global AdS3. 
In each  figure, three constant-$t$ slices are shown. Brown lines are constant $r$, blue 
lines constant $\phi$, red lines where removed conical wedge meets the boundary, 
and black lines where the constant-$t$ surface meets the boundary. 
In both cases $A\ell = 1.7$, $x_0=-y_h-0.1=-0.90869$, and the three $t$ values are $0, \pm \pi/2$.
(\ref{fig:FastLight1}) 
Accelerating point particle ($j=0$). (\ref{fig:FastLight2}) 
Accelerating and rotating point particle with $j = 0.25$. The brown central 
region denotes the excised region of closed timelike curves.
}
\label{fig:Class1FastLight}
\end{figure}

\subsection{Saturated acceleration}

We may also introduce polar-like coordinates for the saturated phase of acceleration,
\begin{equation}
    \tau~=~At\,,\quad x~=~\cos(\alpha \phi)\,,\quad y~=~-\frac{1}{Ar}\,,
\end{equation}with $\alpha<1$. Then, the line element of the Class I rotating C-metric reads
\begin{subequations}
    \begin{align}
        d s^2&~=~\frac{1}{\Omega^2(r,\phi)}\lrb{-\lr{d t+Sd\phi}^2+d r^2+\alpha^2  r^2d\phi^2}\,,\\
        \Omega(r,\phi)&~=~1+A r\cos(\alpha \phi)\,,
    \end{align}
\end{subequations}
where $S/\alpha=j/A$. This line element is conformal to a rotating point particle in 
Minkowski spacetime~\cite[(2.4)]{LoukoKepler}. The tension of the domain wall is given by
\begin{equation}
    \sigma~=~\frac{A}{4\pi}\sin(\alpha\pi)\,.
\end{equation}

For $\alpha<1/2$, the conformal boundary is completely hidden; 
however, for $1/2<\alpha<1$, the conformal boundary is located at
\begin{equation}
    r_{\mrm{C}}~=~-\frac{1}{A\cos(\alpha\phi)}\,,
\end{equation}for $\alpha|\phi|\in(\frac\pi2,\pi)$.

\subsection{Rapid acceleration}\label{sec: rapid pp}

\begin{figure}
\centering
    \begin{subfigure}[b]{0.49\textwidth}
    \centering
    \includegraphics[width=\textwidth]{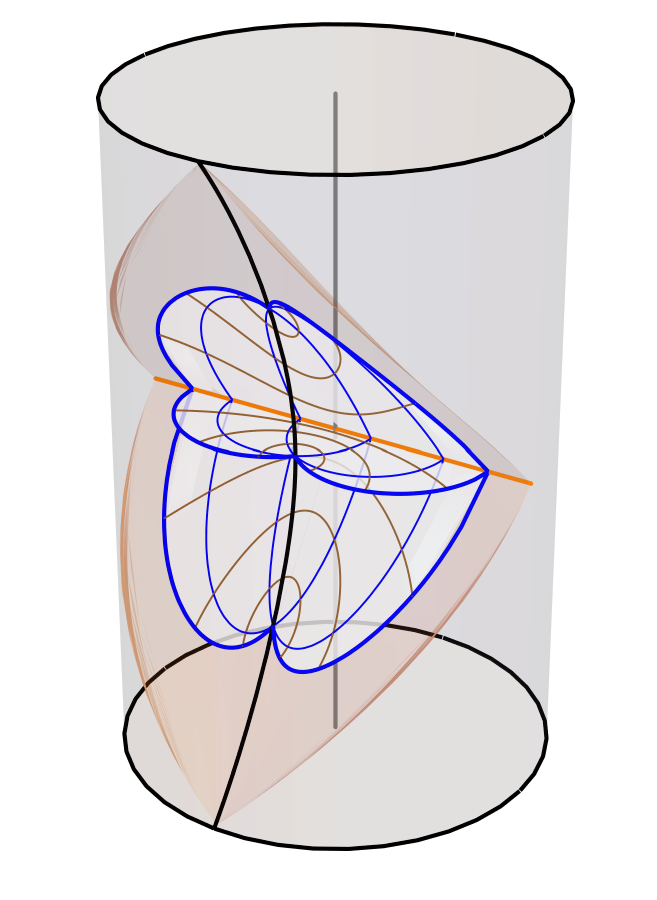}
    \caption{\label{fig:FastHeavy1}}
    \end{subfigure}
    \begin{subfigure}[b]{0.50\textwidth}
    \centering
    \includegraphics[width=\textwidth]{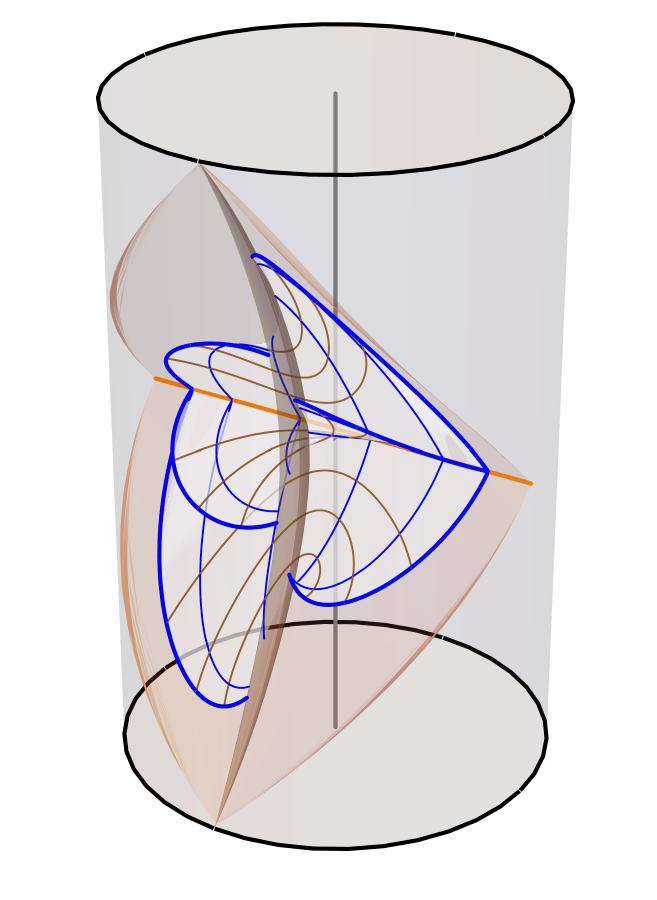}
    \caption{\label{fig:FastHeavy2}}
    \end{subfigure}
\caption{Heavy, rapidly accelerating point particles in AdS mapped onto global AdS3. In each  figure, three constant-$t$ slices are shown. Brown lines are constant $y (\rho)$, blue lines constant $x (\phi)$, red lines where removed conical wedge meets the boundary, and black lines where the constant-$t$ surface meets the boundary.
In both cases $A\ell = 1.7$, $x_0=-y_h+0.1=-0.70869$, and the three $t$ values are $0, \pm \pi/2$
(\ref{fig:FastHeavy1}) Accelerating point particle ($j=0$). (\ref{fig:FastHeavy2}) Accelerating and rotating point particle ($j = 0.25$) with brown central region denoting the excised region of closed timelike curves.\label{fig:Class1FastHeavy}}
\end{figure}

The structure of rapidly accelerating point particles is more interesting.
In the rapid phase of acceleration, the point particle is no longer static with respect to
global coordinates and instead follows a hyperbolic trajectory in the AdS3 cylinder. 
The resulting spacetime is characterised by the presence of an acceleration horizon
at $|y| = |y_h| = \sqrt{ 1 - 1/A^2\ell^2}$, 
and neither the canonical nor the polar patches cover the full spacetime. See appendix 
\ref{globalstructure} for a full discussion. For
the polar coordinates we introduce
\begin{equation}
    \tau~=~\frac{m^2\mc{A}(1+2a^2)t}{\sqrt{1+a^2}}\,,\quad x~=~\cos(m\sqrt{1+a^2}\,\phi)\,,\quad y~=~-\frac{\sqrt{1+2a^2}}{\mc{A}\sqrt{\rho^2+a^2\ell_m^2m^2}}\,,
\end{equation}
where we have set $A=\mc{A}m$, $\ell_m^2=\ell^2/(\mc{A}^2\ell^2m^2-1)$, 
$j^2=a^2\mc{A}^2\ell^2_mm^2/(1+a^2)$, and require $m\sqrt{1+a^2}<1$. 
The line element becomes
\begin{subequations}
    \begin{align}\label{eq: class I rot rapid}
        d s^2&~=~\frac{1}{\Omega^2(\rho,\phi)}\lrb{-g(\rho)d t^2+\frac{d\rho^2}{g(\rho)}+\rho^2\lr{d\phi+\frac{J}{2}\lr{\frac{1}{\rho^2}-\frac{1}{\rho_A^2}}d t}^2}\,,\\
        \Omega(\rho,\phi)&~=~1+\frac{\mc{A}}{\sqrt{1+2a^2}}\sqrt{\rho^2+a^2\ell_m^2m^2}\cos(m\sqrt{1+a^2}\,\phi)\,,\\
        g(\rho)&~=~M-\frac{\rho^2}{\ell_m^2}+\frac{J^2}{4\rho^2}\,,
    \end{align}
\end{subequations}
where $M=m^2$, $J=2a\ell_mm^2\sqrt{1+a^2}$, and 
$\rho_A^2=m^2\ell_m^2(1+a^2)$ denotes the position of the acceleration horizon. 
The tension of the domain wall is given by
\begin{equation}
    \sigma~=~\frac{\mc{A}m}{4\pi}\sin(m\pi\sqrt{1+a^2})\,.
\end{equation}

The line element~\eqref{eq: class I rot rapid} is conformal to that of a rotating point 
particle in de Sitter spacetime~\cite{KdS3D1,KdS3D,MassEntropydeSitter} and the 
position of the acceleration horizon can be interpreted as a cosmological 
horizon in de Sitter. By noting that the coefficient of $d\phi d t$ in~\eqref{eq: class I rot rapid} 
vanishes at $\rho=\rho_A$, we see that these coordinates are in a frame co-rotating 
with the acceleration horizon.

In addition to the acceleration horizon, the conformal boundary is present within these 
coordinates if $1/2<m\sqrt{1+a^2}<1$. There is a conformal boundary at 
\begin{equation}
\label{confbdrho}
    \rho_{\mrm{C}}^2~=~\frac{1+2a^2}{\mc{A}^2\cos^2(m\sqrt{1+a^2}\,\phi)}-a^2\ell_m^2m^2\,,
\end{equation}
for $m\sqrt{1+a^2}|\phi|\in\lr{\frac\pi2,\arccos(-\frac{\sqrt{1+2a^2}}{a\mc{A}\ell_mm})}$. 
However, the lack of a boundary in the polar patch is no longer necessarily a
coordinate shortcoming, for there is a key new possibility for rapidly accelerating 
particles, first identified in \cite{Accin3D}. Recall that the size of the angular deficit 
defining the point particle determines its mass parameter (as well as the tension
of the string pulling it). The curves of constant angle curve away from the point particle
into the bulk. The closer to the boundary the particle, and the greater the deficit angle,
the further round these constant $\phi$ lines curve. For rapid acceleration, the
possibility arises that these constant $\phi$ lines curve so strongly that they arc back
to the acceleration horizon rather than reaching the boundary. 
This limit, $4\pi\sigma\ell>1$, can be determined either by setting $\rho = \rho_A $ 
in \eqref{confbdrho}, or by rearranging $-y_h < x_0$.
The particle
has become so ``heavy'' that it has compactified the space around it
and represents a distinct
global structure.
In the terminology of
\cite{Accin3D}, we refer to these strongly accelerated solutions as \textit{heavy} 
point particles, with \textit{light} point particles satisfying $4\pi\sigma\ell<1$.

\section{Accelerating and rotating black holes}
\label{Sec:BTZJ}

\begin{figure}
\centering
    \begin{subfigure}[b]{0.45\textwidth}
    \centering
    \includegraphics[width=\textwidth]{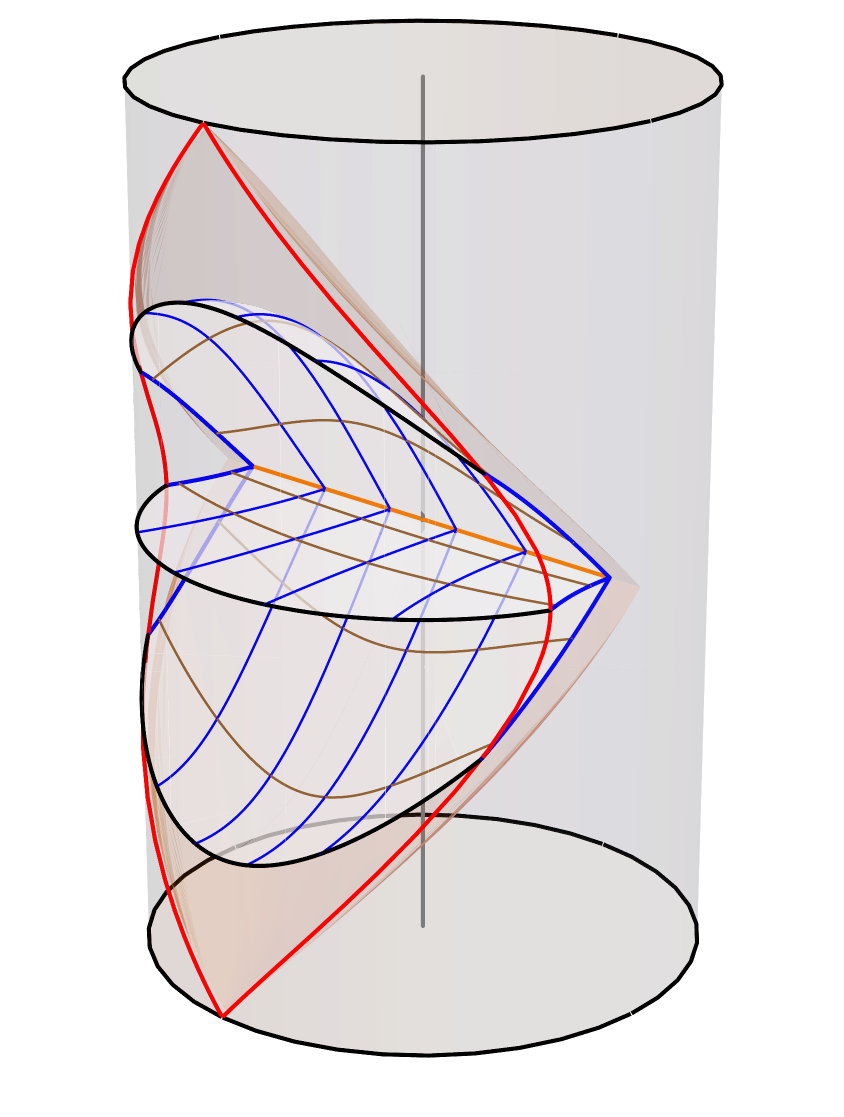}
    \caption{\label{fig:C2Wall2}}
    \end{subfigure}
    \begin{subfigure}[b]{0.45\textwidth}
    \centering
    \includegraphics[width=\textwidth]{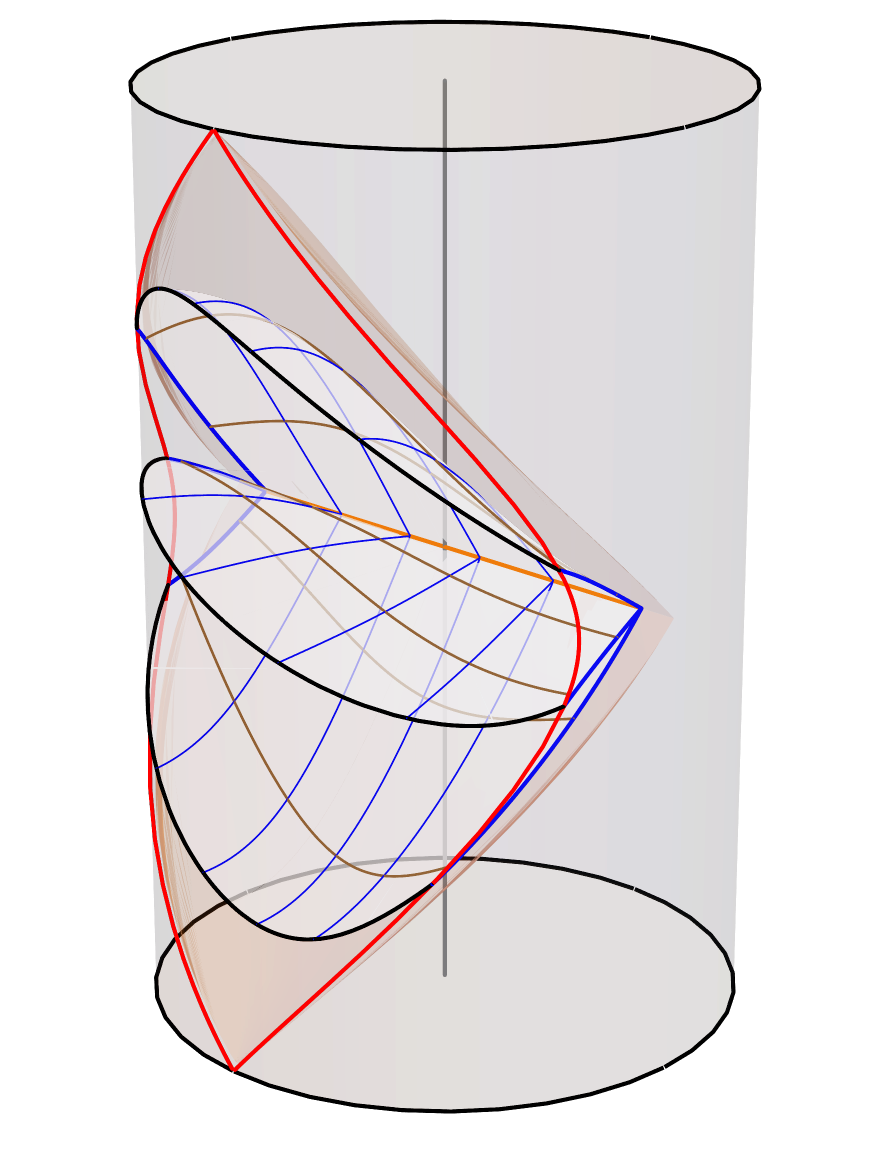}
    \caption{\label{fig:C2Wall3}}
    \end{subfigure}
\caption{Black holes accelerated by a string in AdS mapped onto global AdS3. 
In each figure, three constant-$t$ slices are shown. 
Brown lines are constant $y (\rho)$, blue lines constant $x (\phi)$, red lines indicate
where the string meets the boundary, and black lines are where the constant-$t$ 
surfaces meet the boundary. The orange line denotes the compact black-hole horizon. In both cases $A\ell = 0.5$, $x_0=-1.8$, and the three $t$ values are $0, \pm \pi/4$.
(\ref{fig:C2Wall2}) Accelerating non-rotating black hole ($j=0$). 
(\ref{fig:C2Wall3}) Accelerating and rotating black hole ($j=0.2$).
\label{fig:C2Wall}}
\end{figure}

The family of black holes in AdS3 is described by Class II metrics, and consists of a subset of
the Rindler wedge in AdS3, with identifications that introduce acceleration and/or rotation.
In canonical coordinates, two mirror patches are identified across $x = 1$ (or $x=-1$),
excising the exterior, $|x|>|x_0|$, part of the spacetime. For $A>0$, the positive $x_0$ 
choice corresponds to {\it negative} tension struts, whereas the negative choice corresponds to {\it positive}
tension strings. In addition, since $y<x$, for positive $x_0$ we have two possibilities, 
depending on which of $x_0$ or $y_h = \sqrt{1 + 1/A^2\ell^2}$ is the larger. If $x_0<y_h$, 
then our range of $y$ is $y\in[-y_h, x]$, for all $x\in[1,x_0]$, and there is only one horizon,
that of the black hole, at $y = -y_h$. However, if $x_0>y_h$, then for $x\in (y_h, x_0)$,
$y$ is bounded above by $y_h$ and not $x$. This means there are {\it two} segments of horizon:
$\{ y=-y_h, x\}$ and $\{ y = y_h, x\in(y_h,x_0]\}$, the second horizon being interpreted as an
acceleration horizon. These two cases are referred to as Slow and Rapid accelerating BTZ
black holes. As before, rotation is introduced by adding a time-shift before identifying (see appendix \ref{globalstructure}).

\begin{figure}
\centering
    \begin{subfigure}[b]{0.49\textwidth}
    \centering
    \includegraphics[width=\textwidth]{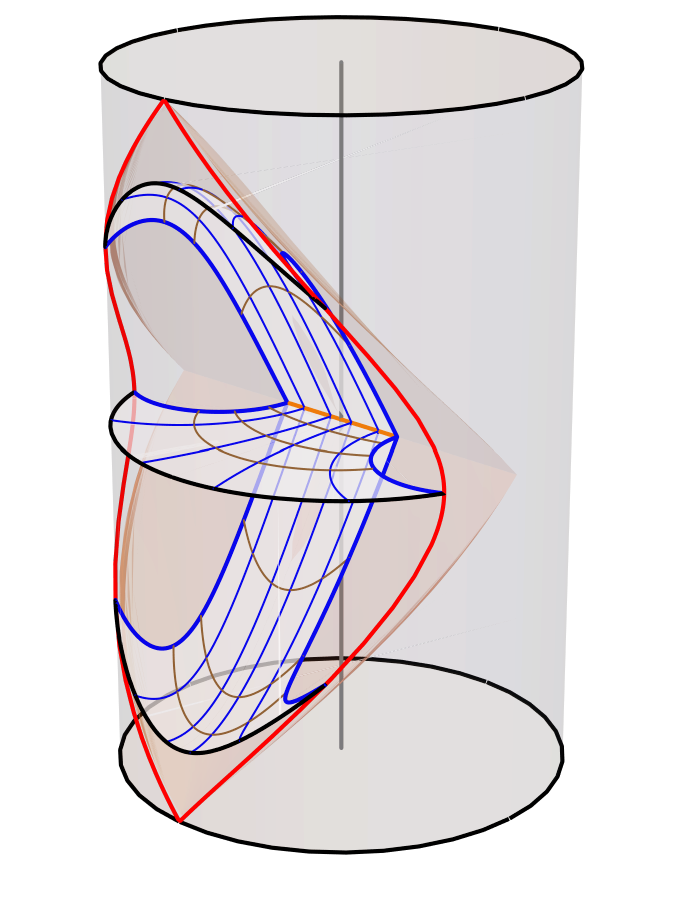}
    \caption{\label{fig:C2StrutSlow2}}
    \end{subfigure}
    \begin{subfigure}[b]{0.49\textwidth}
    \centering
    \includegraphics[width=\textwidth]{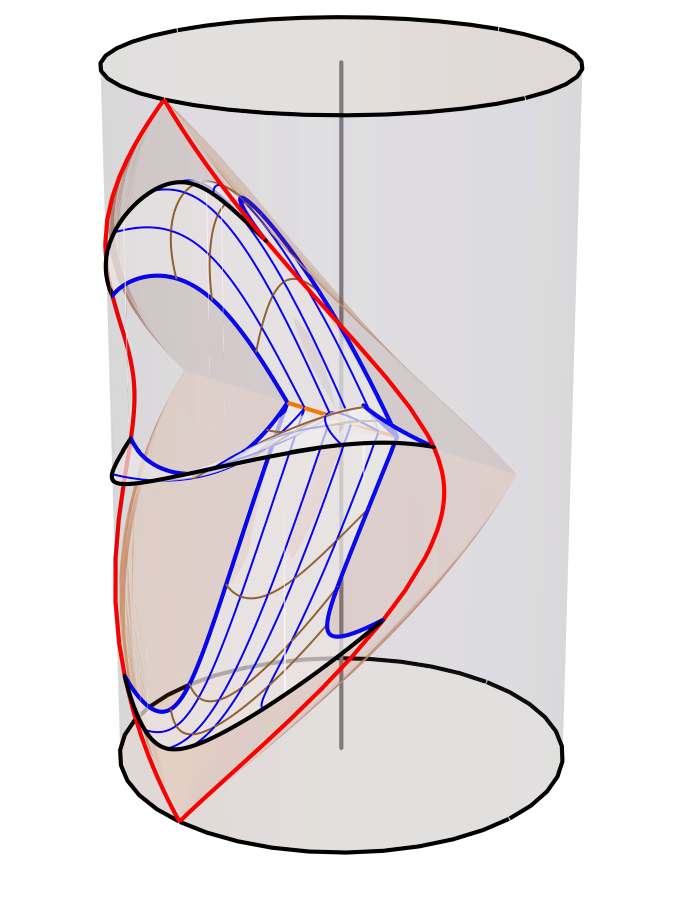}
    \caption{\label{fig:C2StrutSlow3}}
    \end{subfigure}
\caption{Slowly accelerating black holes in AdS mapped onto global AdS3 
with a strut providing the acceleration. 
In each figure, three constant-$t$ slices are shown. 
Brown lines are constant $y$ ($\rho$), blue lines constant $x$ ($\phi$), 
red lines indicate where the strut meets the boundary, 
and black lines where the constant-$t$ surfaces meet the boundary. 
In both cases $A\ell = 0.5$, $x_0=1.8<y_h$, and the three $t$ values are $0, \pm\pi/4$.
The orange line denotes the compact black-hole horizon. 
(\ref{fig:C2StrutSlow2}) Non-rotating slowly accelerating black hole ($j=0$). 
(\ref{fig:C2StrutSlow3}) Accelerating and rotating black hole with $j = 0.2$.
\label{fig:C2StrutSlow}}
\end{figure}

These possibilities are better interpreted via the polar-like coordinates
\begin{equation}\label{eq: class 2 negative tension coords}
    \tau~=~\frac{m^2\mc{A}t}{\sqrt{1+a^2}}\,,\quad x~=~\mp \cosh(m\sqrt{1+a^2}\,\phi)\,,\quad y~=~-\frac{1}{\mc{A}\sqrt{\rho^2-\rho_-^2}}\,,
\end{equation}where $t\in\RR$, $\rho>\rho_-$, $\phi\in(-\pi,\pi)$, and we have set $A=\mc{A}m$, $\ell_m^2=\ell^2/(1+\mc{A}^2\ell^2m^2)$, $\rho_-^2=a^2\ell_m^2 m^2$, and $j^2=a^2\mc{A}^2\ell_m^2m^2/(1+a^2)$. 
We note that the negative sign of $x$ corresponds to a positive-tension solution,
and the positive sign to a negative tension solution.
The line element of the Class II rotating C-metric reads
\begin{subequations}\label{eq:ds C2 pos ten}
    \begin{align}
        d s^2&~=~\frac{1}{\Omega^2(\rho,\phi)}\lrb{-g(\rho)d t^2+\frac{d \rho^2}{g(\rho)}+\rho^2\lr{d\phi+\frac{J}{2}\lr{\frac{1}{\rho^2}-\frac{1}{\rho_+^2}}dt}^2}\,,\\\label{eq: class 2 Omega pos tension}
        \Omega(\rho,\phi)&~=~1 \mp\mc{A}\sqrt{\rho^2-\rho_-^2}\cosh(m\sqrt{1+a^2}\,\phi)\,,\\
        g(\rho)&~=~\frac{\rho^2}{\ell_m^2}-M+\frac{J^2}{4\rho^2}\,,
    \end{align}
\end{subequations}
where $M=(1+2a^2)m^2$, $J=2a\ell_m m^2\sqrt{1+a^2}$,  $\rho_+^2=m^2\ell_m^2(1+a^2)$, and 
$\rho_-^2=a^2m^2\ell_m^2$. Analogously to the BTZ black hole, we refer to $\rho_+$ and $\rho_-$ as the outer and inner horizons respectively. However, the range of the radial coordinate does not include the inner horizon. This is consistent with the fact that $P(y)=0$ (cf. Table~\ref{tab:classes}) has only one root for $y<0$, corresponding to $\rho=\rho_+$. The tension of the domain wall is given by
\begin{equation}
    \sigma~=~ \pm \frac{\mc{A}m}{4\pi}\sinh(m\pi\sqrt{1+a^2})\,.
\end{equation}

\begin{figure}
\centering
    \begin{subfigure}[b]{0.49\textwidth}
    \centering
    \includegraphics[width=\textwidth]{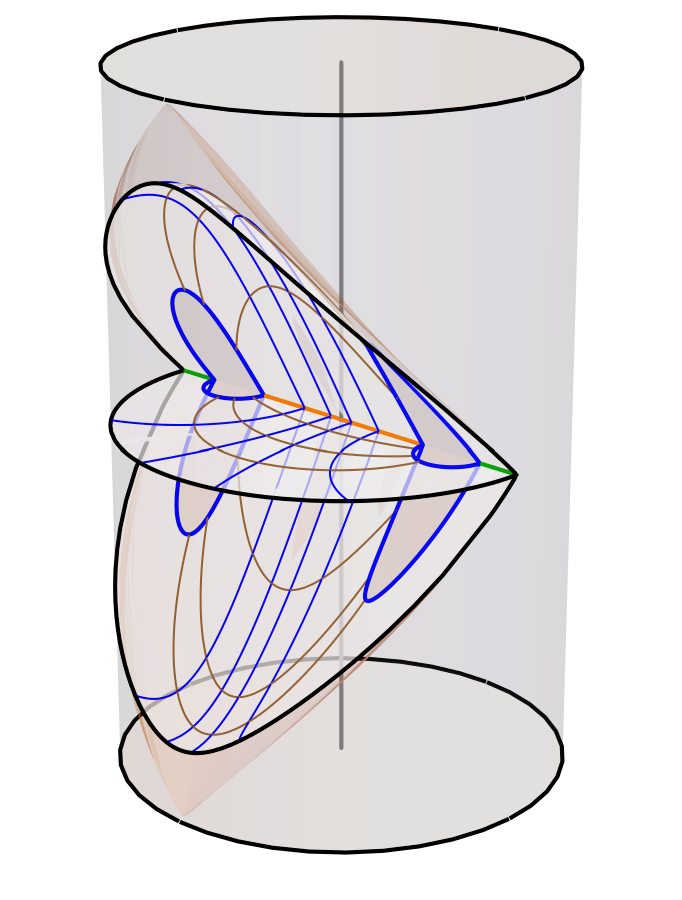}
    \caption{\label{fig:C2StrutFast1}}
    \end{subfigure}
    \begin{subfigure}[b]{0.49\textwidth}
    \centering
    \includegraphics[width=\textwidth]{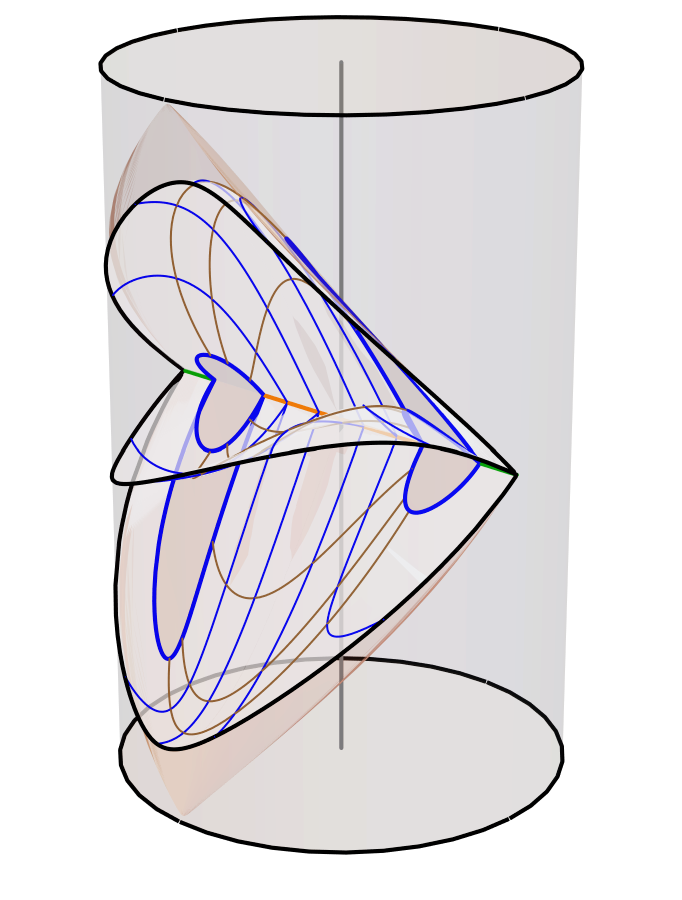}
    \caption{\label{fig:C2StrutFast2}}
    \end{subfigure}
\caption{Rapidly accelerating BTZ black holes in AdS mapped onto global AdS3. 
In each figure, three constant-$t$ slices are shown. Brown lines are constant $y/r$, 
blue lines constant $x/\phi$, and black lines indicate where the constant-$t$ 
surface meets the boundary. 
In both cases $A\ell = 0.5$, $x_0=4>y_h$, and the three $t$ values are $0, \pm\pi/4$.
The orange line denotes the compact black-hole horizon, 
whilst the green lines indicate the non-compact acceleration horizon. 
(\ref{fig:C2StrutFast1})~Accelerating non-rotating black hole ($j=0$). 
(\ref{fig:C2StrutFast2}) Accelerating and rotating black hole with $j=0.2$.
\label{fig:C2StrutRapid}}
\end{figure}

The conformal boundary, corresponding to the AdS boundary, is located at
\begin{equation}
    \rho^2_{\mrm{C}}~=~\frac{1}{\mc{A}^2\cosh^2(m\sqrt{1+a^2}\,\phi)}+\rho_-^2\,.
\end{equation}Imposing that the outer horizon lie fully within the conformal boundary, $\rho_+<\rho_{\mrm{C}}$, we require
\begin{equation}
    \mc{A}\ell m\sinh(m\pi\sqrt{1+a^2})~<~1\,.
\end{equation}

As with slowly accelerating point particles (cf. section~\ref{sec: slow pp}) 
and rapidly accelerating point particles (cf. section~\ref{sec: rapid pp}), 
the coordinates~\eqref{eq: class 2 negative tension coords} with the plus choice 
do not cover the entire spacetime. 
A second patch covering $y>0$, equivalently $\rho<0$, is required. 
The two patches are glued together across $y=0$. 
Within this second patch, the only change to the line element is in the 
conformal factor $\Omega$, which is instead given by~\eqref{eq: class 2 Omega pos tension} 
with the minus choice. The conformal boundary is located at
\begin{equation}
    \rho_{\mrm{C}}~=~-\sqrt{\frac{1}{\mc{A}^2\cosh^2(m\sqrt{1+a^2}\,\phi)}+\rho_-^2}\,.
\end{equation}There is an additional horizon located at $y=1/(\mc{A}\ell_mm)$, equivalently $\rho=-\rho_+$. For $\rho_{\mrm{C}}<-\rho_+$, the horizon is entirely hidden behind the conformal boundary. As in~\cite{Accin3D}, we call this the \textit{slow} phase of acceleration and it is defined by the constraint
\begin{equation}
    \mc{A}\ell m\sinh(m\pi\sqrt{1+a^2})~<~1\,.
\end{equation}
Accelerating and rotating black holes that satisfy $\mc{A}\ell m\sinh(m\pi\sqrt{1+a^2})>1$, 
however, undergo \textit{rapid} acceleration. 
In this rapid phase of acceleration, a non-compact black ``droplet''
\cite{BlackDroplet,BlackDroplet1}, or acceleration horizon, forms.

Finally, as noted in \cite{Accin3D}, within the Class I solutions there is a corner of parameter space that corresponds to an accelerating black hole --- dubbed Class I$_\mrm{C}$. Recall $P = y^2 - y_h^2$
for class I, with $y_h^2 = 1 - 1/A^2\ell^2$.
For $A\ell>1$, $y_h\in(0,1)$ and choosing $x_0>y_h$
allows for an accelerating solution with $y\in[y_h,x]$.
The appearance of the corresponding spacetime is very similar to figure \ref{fig:C2Wall} (see \cite{Accin3D}) so we do not
include a global plot here.

To get the polar system, one uses a similar transformation to 
\eqref{eq: C I slow polar}: 
\be
\label{eq:C1-BTZ}
\tau~=~\frac{m^2\mc{A}t}{\sqrt{1-a^2}}\,,\quad 
x~=~\cos(m\sqrt{1-a^2}\,\phi)\,,\quad 
y~=~\frac{1}{\mc{A}\sqrt{\rho^2+m^2a^2\ell_m^2}}\,,
\end{equation}
where now $A=\mc{A}m$, $\ell_m^2=\ell^2/(\mc{A}^2\ell^2m^2-1)$, and 
$j^2=a^2\mc{A}^2\ell_m^2m^2/(1-a^2)$. 
Defining 
$M=m^2(1-2a^2)$ and $J=2a\ell_mm^2\sqrt{1-a^2}$,
and the functions
\be
\beal
\Omega(\rho,\phi)&~= 1-\mc{A}\sqrt{\rho^2+m^2a^2\ell_m^2}
\cos(m\sqrt{1-a^2}\,\phi)\,,\\
g(\rho) &= - \frac{\rho^2}{\ell_m^2}+M+\frac{J^2}{4\rho^2}\,,
\eeal
\ee
the metric becomes
\be
ds^2 = \frac{1}{\Omega^2}\left [ -g(\rho)dt^2+\frac{d\rho^2}{g(\rho)}
+\rho^2\left ( d\phi+\frac{J}{2}\lr{\frac{1}{\rho^2}+\frac{1}{m^2\ell_m^2(1-a^2)}}dt \right )^2 \right ]\,
\ee
Note, the requirement
that $y_h<y<x$ implies that the range of the $\rho$ coordinate
is now $\rho_C<\rho<\rho_h$, where $\rho_h = m \ell_m \sqrt{1-a^2}$
is the location of the horizon, and 
$$
\rho_{\mrm{C}}^2 = \frac{1}{\mc{A}^2\cos^2(m\sqrt{1-a^2} \phi)}
- m^2 a^2 \ell_m^2
$$
is the location of the conformal boundary.

\section{Class III}
\label{sec:C3}

\begin{figure}
\centering
\includegraphics[width=0.48\textwidth]{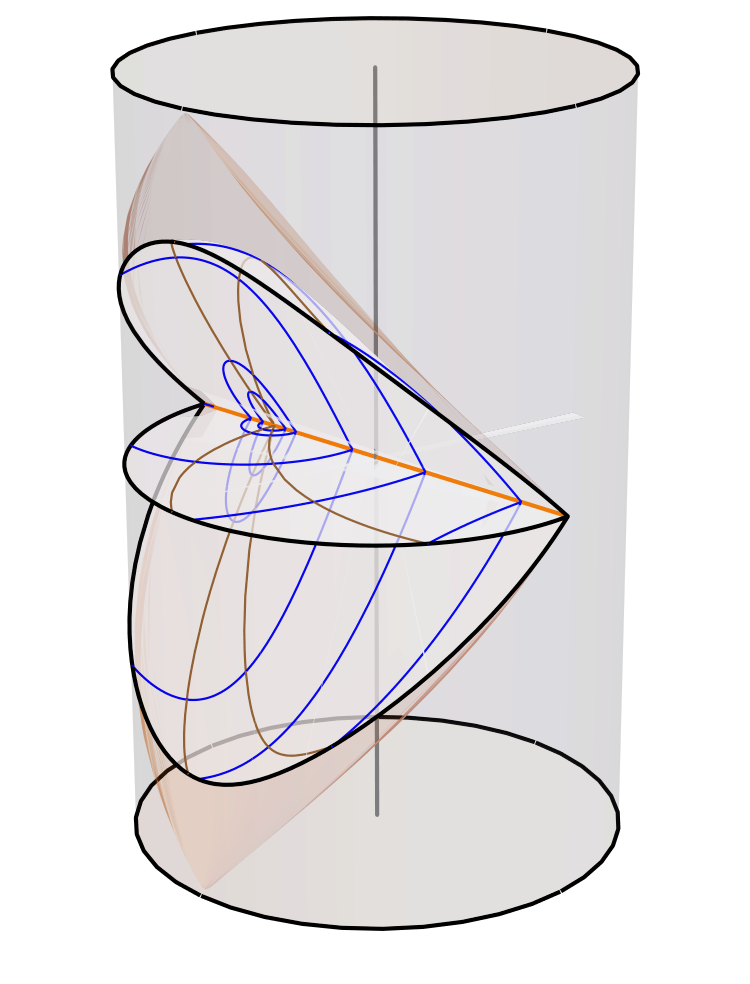}
\includegraphics[width=0.48\textwidth]{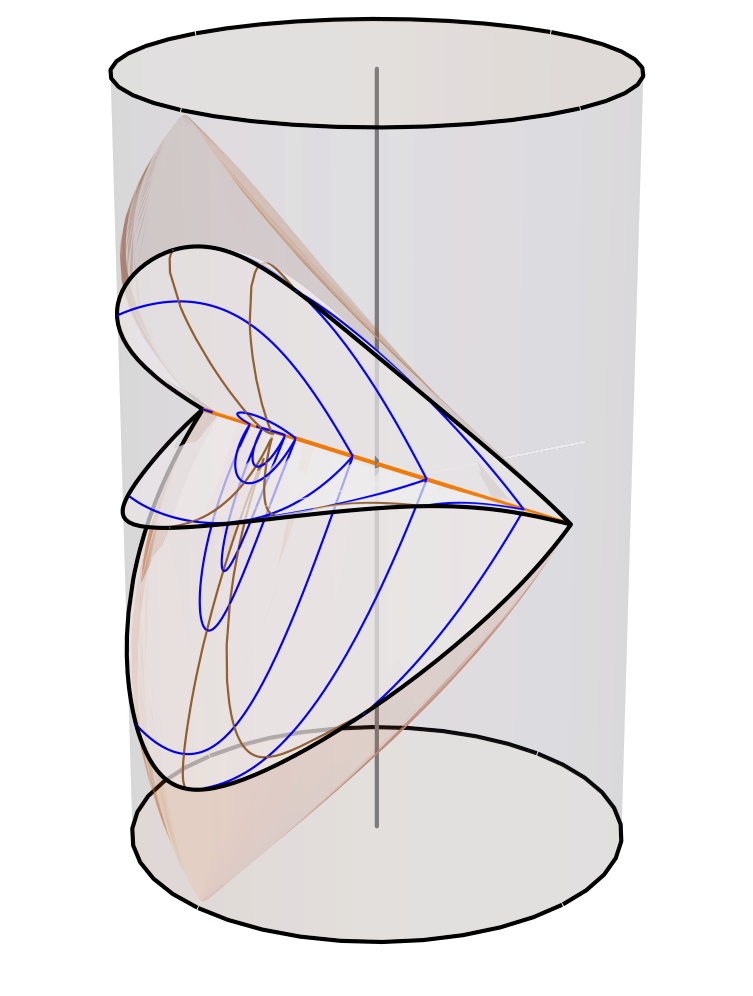}
\caption{The slicing of AdS3 in Class III coordinates with $A\ell = 0.5$, and $j=0.2$ for the ``rotating'' slicing on the right. 
Constant-$x$ lines are in blue, and constant-$y$ in brown. Three constant time slices are shown. The accumulation point on the left side is $x \to \infty$, and $x\to y_h$ on the right hand side.}
\label{fig:C3slices}
\end{figure}

Class III solutions are different in that they have no reflection symmetry, instead the coordinates parametrise 
AdS in an asymmetric manner, see figure~\ref{fig:C3slices}.
Thus, all constant-$x$ surfaces have a distinct induced metric and
we cannot match ``mirror image'' surfaces in the spacetime. 
We therefore require {\it two} copies of the spacetime to match along a constant-$x$ surface, 
analogous to the Randall-Sundrum braneworld picture \cite{Randall:1999ee,Randall:1999vf}, 
with identification across one or two walls.
Without these identifications, adding rotation has no physical meaning, 
as it is simply another relabelling of AdS3.

\begin{figure}
\centering
    \includegraphics[width=0.4\textwidth]{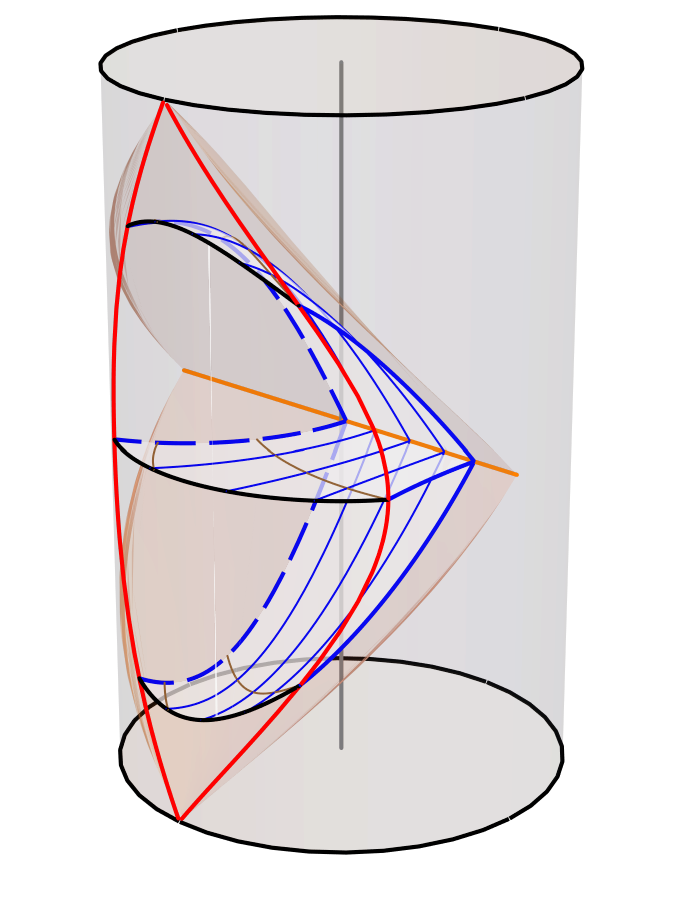} 
    \includegraphics[width=0.1\textwidth]{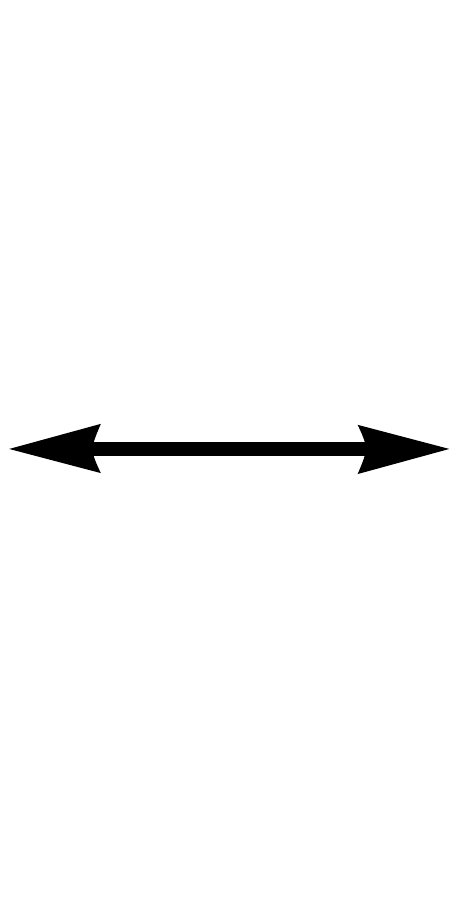} 
    \includegraphics[width=0.4\textwidth]{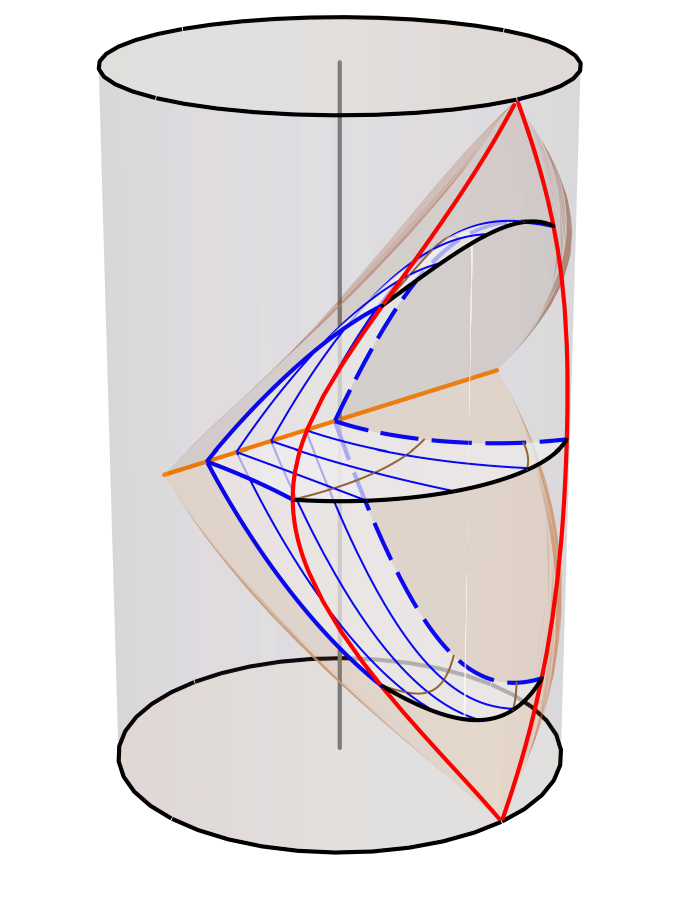}
\caption{The RS1 Class III braneworld set-up without rotation. Two mirror copies of the 
Class III spacetime cut along two constant-$x$ lines are shown, with the identified
constant-$x$ surfaces indicated in thick blue with the positive tension surface indicated
by a solid line and the negative tension surface by a dashed line. 
Thin blue and brown lines indicate constant $x$ and $y$ respectively, and
the thick red lines indicate the limit of the retained spacetime on
the boundary. The thick black lines on the boundary are at constant $\tau$.
The parameters used in these plots are $A\ell = 0.5$, with the two cuts at
$x_\pm = -1, 0.5$.}
\label{fig:C3RS1}
\end{figure}

\begin{figure}
\centering
    \includegraphics[width=0.4\textwidth]{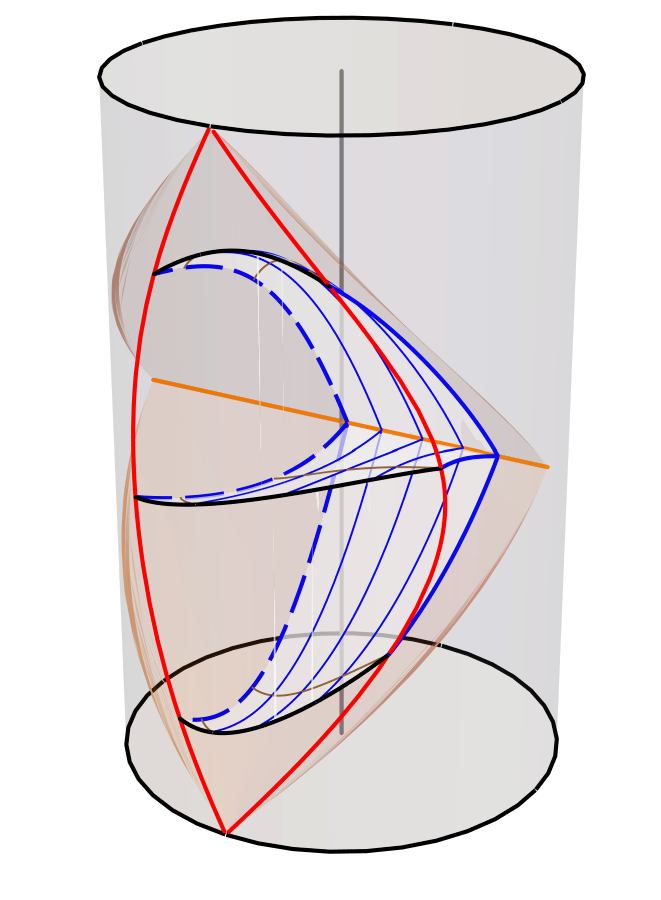}
    \includegraphics[width=0.1\textwidth]{arrowfig.pdf} 
    \includegraphics[width=0.4\textwidth]{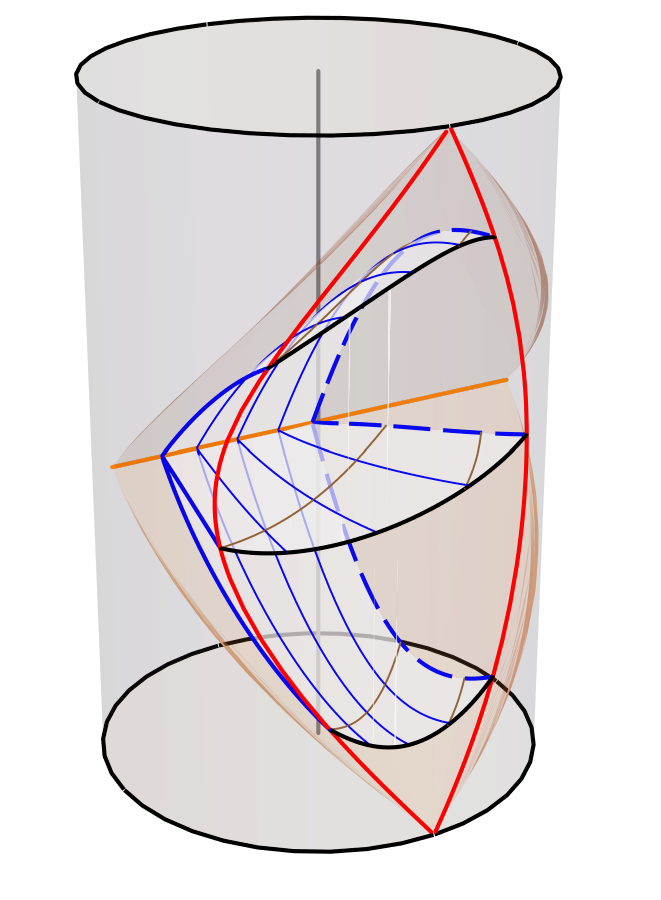}
\caption{The RS1 Class III braneworld set-up with rotation. 
As in figure \ref{fig:C3RS1}, $A\ell = 0.5$, and the two cuts are at
$x_\pm = -1, 0.5$ with $j = 0.2$ additionally. The shift can be clearly seen.}
\label{fig:C3RS1-J}
\end{figure}

For illustration, figures \ref{fig:C3RS1} and \ref{fig:C3RS1-J} show precisely this
Randall-Sundrum-1 (RS1) braneworld set-up without and with rotation respectively.
Two constant-$x$ surfaces are chosen and the two copies identified along these two branes.
Like the RS1 model, one of the branes has positive tension, and the other negative tension. 
The spacetime with rotation is shown in figure \ref{fig:C3RS1-J}, where the time-shift in identification is shown.
Crucially, note that we no longer need to remove a chronology violating region.

\section{Conclusions}
\label{sec:wrap}

In summary, we have classified all accelerating, rotating solutions to Einstein gravity with a negative cosmological constant in three dimensions. The key to including rotation is to note that the 3D spacetimes are segments of global anti-de Sitter space with nontrivial identifications, with rotation introduced via a time-shift across 
the identifying surface, resulting in a helical identification. The accelerating rotating spacetimes share
many features with their non-accelerating cousins such as the existence of a chronology limit surface and evidence for inner Cauchy horizons in the black hole solutions.
The main difference with these spacetimes is the presence of the string or strut, that extends to the AdS boundary (when the boundary is present in the solution). This translates into a defect in the boundary geometry. Following \cite{RuthAccBH}, we anticipate an enhancement of the thermodynamic phase space of these solutions that we leave for future work \cite{inprep}.

\acknowledgments \label{sec:: acknowledgements}

CRDB thanks Jorma Louko for helpful discussions. 
The work of CRDB was supported by United Kingdom Research and Innovation 
Engineering and Physical Sciences Research Council (EPSRC) [grant number EP/W524402/1]. 
RG is supported in part by STFC (ST/X000753/1) and in part by the Perimeter 
Institute for Theoretical Physics.
Research at Perimeter Institute is supported in part by the Government of Canada
through the Department of Innovation, Science and Economic Development and by
the Province of Ontario through the Ministry of Colleges and Universities.
RG also acknowledges support from the Simons Center for Geometry and Physics, 
Stony Brook University during the programs 
{\it Geometry and Convergence in Mathematical General Relativity}
and {\it 50 Years of the Black Hole Information Paradox}
at which some of the research for this paper was performed.
RBM also acknowledges support from the Simons Center for Geometry and Physics, 
Stony Brook University and the program {\it 50 Years of the Black Hole Information Paradox}, and support from the Natural Sciences and Engineering Research Council of Canada.

For the purpose of open access, the authors have applied a CC BY 
public copyright licence to any Author Accepted Manuscript version arising. 
No new data were created during this study.

\appendix

\section{Global Structure}\label{globalstructure}

In order to visualise the global structure of the accelerating 
solutions, we follow the method outlined in the appendix 
of \cite{Accin3D}. 
For the purpose of plotting, the canonical coordinates prove to be
the most suitable as they have the most comprehensive coverage of the
physical spacetime, though care must be taken when patching together the
two copies along the zero tension $|x|=1$ surface (that is neatly 
accommodated in the angular coordinate). 
Recall the canonical rotating metric \eqref{transform-to-t}, 
\be
ds^2 = \frac{1}{A^2 (x-y)^2} \left [ - P(y) \left ( d\tau - j \frac{dx}{Q(x)} \right )^2 
+ \frac{dy^2}{P(y)} + \frac{dx^2}{Q^2(x)} \right]
\ee
where $P$ and $Q$ are given in table \ref{table:ClassParameters}. We 
use the transformation rules of \cite{Accin3D} for
$d\bar{t} = d\tau - j dx/Q$, then replace $t$ accordingly.
Here we briefly recap this plotting process.

Recall that global AdS$_3$, \eqref{eq: universal covering spacetime}, can be plotted
as a cylinder by conformally transforming the spatial sections (here parametrised by
$X_G=R_G\cos\Phi, Y_G = R_G\sin\Phi$) to the Poincar\'e disc. There are several options for this compactification; here we use the transformation
\be
T_{\text{CYL}} = \frac{T_G}{\ell} \;,\;\;
X_{\text{CYL}} = \frac{X_G}{\ell + \sqrt{\ell^2 + X_G^2 + Y_G^2}} \;,\;\;
Y_{\text{CYL}} = \frac{Y_G}{\ell + \sqrt{\ell^2 + X_G^2 + Y_G^2}} 
\ee

To obtain the plots of the various Classes of solution, we note that
AdS$_3$ can also be represented as a hyperboloid embedding
$X_1^2 + X_2^2 - X_3^2 - X_4^2 = - \ell^2$ in a
4D spacetime with signature $+,+,-,-$, where in terms of the global
AdS coordinates $(T_G, X_G, Y_G)$
\be
\beal
X_1 &= X_G = R_G \cos \Theta_G \quad &
X_3 &= \sqrt{\ell^2 + X_G^2 + Y_G^2} \cos (T_G/\ell) \\
X_2 &= Y_G = R_G \sin \Theta_G &
X_4 &= \sqrt{\ell^2 + X_G^2 + Y_G^2} \sin (T_G/\ell)
\eeal
\ee
which is easily inverted to obtain $(T_G,X_G,Y_G)$.

One then finds a map that transforms the canonical
coordinates in the various classes to this hyperboloidal embedding space, thus providing a parametric map between
these coordinates and the cylindrical plotting ones.
Note that this map is between the non-rotating Class I-III
solutions; this is all that is required since rotation is 
obtained by adding a shift $dt = d\tau - j dx/Q$ (from
\eqref{transform-to-t}), which allows the nontrivial 
identification of constant time surfaces in the bulk cylinder.

The transformation between the canonical coordinates and the covering space
can be encoded in the following {\it Master Equation:}
\be
\beal
Y_a &= \frac{\sqrt{-\epsilon_0 \epsilon_2} Q(x)}{\Omega(x,y)}  \quad &
Y_c &= \frac{A\ell \sqrt{\epsilon_0 \epsilon_2}}{ y_h \Omega(x,y)} 
\left ( y_h^2 x + \epsilon_0 \epsilon_2 y \right ) \\
Y_b &= \frac{\sqrt{P(y)}}{y_h \Omega(x,y)}\cosh \left ( y_h t \right ) \quad & 
Y_4 &= \frac{\sqrt{P(y)}}{y_h\Omega(x,y)}\sinh \left ( y_h t \right ) 
\eeal
\label{mastertransform}
\ee
where
\be
Y_a^2 + Y_b^2 + Y_c^2 - Y_4^2 = -\ell^2
\ee
and
\be
y_h^2 = \epsilon_2 \left ( \frac{1}{A^2 \ell^2} - \epsilon_0 \right )
\ee
Here, the $\epsilon_i$ values are given in table \ref{tab:classes}, and the metric functions in \eqref{eq:canonmetfns}. 
Note that the appropriate signature for the covering space is achieved by one of the $Y_{a,b,c}$
being imaginary, as can clearly be seen by the appearance of the root of both 
$\epsilon_0\epsilon_2$ and $-\epsilon_0\epsilon_2$.

We now briefly state the transformation for each of the various classes, 
noting the parameter ranges / constraints for the various cases 
as one of the main niggles in plotting
is ensuring the right range of parameters is taken!

\subsection{Class I}

The Class I metrics have $\epsilon_0=-\epsilon_2 = 1$, thus
\be
Q^2(x) = 1-x^2 \quad , \qquad P(y) = \frac{1}{A^2\ell^2} - (1-y^2) = y^2 - y_h^2
\ee
where the magnitude of $A\ell$ determines whether or not there is an 
acceleration horizon.

\subsubsection{Slow Acceleration}

In slow acceleration, $A\ell<1$, $\epsilon_0\epsilon_2=-1$, and $y_h$ is imaginary, giving
\be
\beal
X_1 &= Y_a = \frac{Q(x)}{\Omega(x,y)}&
X_3 &= iY_b = \frac{\sqrt{P(y)}}{\alpha \Omega(x,y)} \cos \left ( \alpha t \right )\\
X_2 &= Y_c = \frac{A\ell\left ( \alpha^2 x + y \right )}{\alpha \Omega(x,y)}  \quad &
X_4 &= Y_4 = \frac{\sqrt{P(y)}}{\alpha \Omega(x,y)}\sin \left ( \alpha t \right )
\eeal
\ee
where $\alpha^2 = -y_h^2$. In this case, the $(x,y)$ space maps directly onto
the $(X_{\text{CYL}},Y_{\text{CYL}})$ constant global time disc:
\be
T_{\text{CYL}} = \alpha t \;\;\;,\quad
X_{\text{CYL}} = \frac{\alpha Q}{\left (\ell\alpha\Omega + \sqrt{P}\right )}\;\;\;,\quad
Y_{\text{CYL}} = \frac{A\ell \left (\alpha^2 x + y\right )}{\left (\ell\alpha\Omega + \sqrt{P}\right )} 
\ee
For the full (non-rotating) spacetime
we take a mirror image and glue along $x=1$ and $x=x_0$. 
Constant $t$ surfaces coincide with constant $T_{\text{CYL}}$
surfaces, and we have the plot in figure \ref{fig:1Slow}.
For the rotating spacetime, we substitute $t = \tau - j\, 
\text{arccos} (x)$ in the primary patch, but in the mirror patch we must replace
$t = \tau + j\, \text{arccos} (x)$. 
These do not alter the three dimensional patches of each
in the cylinder, however do alter the identification across the surface,
which now has a time-shift:
\be
\left (T_{\text{CYL}},X_{\text{CYL}}
(x_0,y),Y_{\text{CYL}}(x_0,y)\right )\sim
\left ( T_{\text{CYL}} - 2 j\, \arccos x_0 ,X_{\text{CYL}}
(x_0,y),Y_{\text{CYL}}(x_0,y)\right )
\ee 
Constant $\tau$ surfaces are also seen to have a nontrivial
$T_{\text{CYL}}$ behaviour. 

\subsubsection{Rapid Acceleration}

A rapidly accelerating point particle is one for which $A\ell>1$, which now gives rise to a
zero of $P$ at $y_h^2 = 1 - 1/A^2\ell^2$, interpreted as an acceleration horizon. Given that
$y<-y_h$ for the primary patch, there are now two possibilities for the wall at $x_0$. If
$x_0<-y_h$, then some of the boundary is included in the spacetime for all $t$; but
if $x_0>-y_h$, then $-y_h$ is the upper limit of $y$ for all $x$, thus the wall extends out to
the acceleration horizon and the boundary is excluded. For the nonrotating cases, these two
situations correspond to smaller or greater deficit angles for the point particle respectively, hence the
monikers ``light'' and ``heavy''.

The map from the Class I - Rapid to the covering space of AdS$_3$ is read off as
\be
\beal
X_1 &= Y_a = \frac{Q(x)}{A(x-y)} \quad &
X_3 &= \frac{Y_c}{i} = \frac{\ell }{ y_h (x-y)} \left ( y_h^2 x - y \right ) \\
X_2 &= Y_b = \frac{\sqrt{P(y)}}{A y_h (x-y)}\cosh \left ( y_h t \right )& 
X_4 &= Y_4 = \frac{\sqrt{P(y)}}{A y_h(x-y)}\sinh \left ( y_h t \right )
\eeal
\label{class1tocover}
\ee
The direct map between the canonical coordinates and the plotting coordinates now no longer has a clean
form, and is best plotted parametrically. To recap, for the coordinate ranges (primary patches) we have:
\be
\beal
\text{Light:} &\quad x \in [x_0,1] \; , \qquad 
y \in (-\infty, \min [x,-y_h ] \bigr ] \\
\text{Heavy:} & \quad x \in [x_0,1] \; , \qquad
y \in (-\infty, -y_h ] 
\eeal
\ee

\subsection{Class II}

For Class II spacetimes, the transformation looks schematically the same as Class I - Rapid
with the proviso that the metric functions, and $y_h$, are different:
\be
Q^2(x) = x^2 -1 \quad , \qquad P(y) = \frac{1}{A^2\ell^2} + (1-y^2) = y_h^2 - y^2
\ee
For the sake of completeness, the embedding functions are:
\be
\beal
X_1 &= Y_a = \frac{Q(x)}{A(x-y)}  &
X_3 &= \frac{Y_c}{i} = \frac{\ell }{ y_h (x-y)} \left ( y_h^2 x - y \right )\\
X_2 &= Y_b = \frac{\sqrt{P(y)}}{A y_h (x-y)}\cosh \left ( y_h t \right ) \quad& 
X_4 &= Y_4 = \frac{\sqrt{P(y)}}{A y_h(x-y)}\sinh \left ( y_h t \right ) 
\eeal
\ee
Whereas for Class I, $x$ could continuously transition from positive to negative values, 
for Class II, $x$ must maintain the same sign throughout the primary and mirror patch 
(which are identified either at $x=1$ or $x=-1$). The choice of the sign of $x$ corresponds
to the sign of the wall tension, with $x>1$ patches having a negative tension, and $x<-1$
patches a positive tension wall. In addition, depending on the values of $x_0$ and $y_h$, 
there is the possibility of having both event and acceleration horizons:
\be
\beal
\text{Wall:} &\quad x \in [x_0,-1] \; , & \;\;\; 
& y \in[-y_h, x] &  -y_h &< x_0 < -1 \\
\text{Strut-Slow:} & \quad x \in [1,x_0] \; , & \;\;\; 
& y \in (-\infty, -y_h ] &  x_0  &< y_h  \\
\text{Strut-Rapid:} &\quad x \in [1,x_0] \; , & \;\;\;  
& y \in (-\infty, \min [x,y_h ] \bigr ] &  y_h &< x_0 
\eeal
\ee
As before, for the rotating spacetime, we substitute $t = \tau - j\, 
\text{arccosh} (x)$ in the primary patch, and
$t = \tau + j\, \text{arccosh} (x)$ in the mirror patch.

\subsection{Class III}

For Class III spacetimes, the metric functions are
\be
Q^2(x) = x^2 +1 \quad , \qquad P(y) = \frac{1}{A^2\ell^2} - (1+y^2) = y_h^2 - y^2
\ee
which require $A\ell<1$.
and the embedding functions are:
\be
\beal
X_1 &= Y_c = \frac{\ell }{ y_h (x-y)} \left ( y_h^2 x - y \right )  &
X_3 &= \frac{Y_a}{i} = \frac{Q(x)}{A(x-y)} \\
X_2 &= Y_b = \frac{\sqrt{P(y)}}{A y_h (x-y)}\cosh \left ( y_h t \right ) \quad& 
X_4 &= Y_4 = \frac{\sqrt{P(y)}}{A y_h(x-y)}\sinh \left ( y_h t \right ) 
\eeal
\ee
And to add rotation, we substitute $t = \tau - j\, 
\text{arcsinh} (x)$ in the primary patch, and
$t = \tau + j\, \text{arcsinh} (x)$ in the mirror patch.

The key difference in Class III metrics is that the $\{x,y\}$ patch now covers the full
ADS3 spacetime. Additionally, due to the conformal factor, the metrics on an $x=$const.\
slice are all distinct, thus there are no ``mirror image'' slices along which it is possible to
identify. In order to get a solution distinct from AdS3 we must therefore take {\it two}
copies of the Class III solution, cut at $x=x_0$ in each, and identify. This will give a 
noncompact ``Randall-Sundrum 2'' type of solution \cite{Randall:1999vf}.
Taking two cuts at $x_0$ and $x_1$ in each spacetime and identifying on the other hand 
gives a compact spacetime with two ``end of the world'' branes, reminiscent of 
the ``Randall-Sundrum 1'' braneworld solution \cite{Randall:1999ee}. 
All options contain a conformal horizon.

\bibliographystyle{JHEP}
\bibliography{zz_bibliography.bib}
\end{document}